\documentclass[aps,prl,twocolumn,superscriptaddress]{revtex4}
\usepackage{amsmath,amssymb}
\usepackage{graphics,graphicx}
\usepackage{dcolumn,bm}
\usepackage{psfrag}
\usepackage{color}
\usepackage{appendix}
\usepackage{multirow}
\usepackage{float}

\newcommand{\cu}
{\affiliation{Department of Physics, University of Calcutta, 92 Acharya Prafulla Chandra Road, Kolkata 700009}}

\newcommand{\srm}
{\affiliation{
Department of Physics, SRM University - AP, Amaravati, Andhra Pradesh - 522240, India}}

\newcommand{\imsc}
{\affiliation{The Institute of Mathematical Sciences, CIT Campus, Taramani, Chennai 600113, India.}}

\begin{document}






\title{Diffusive-to-Ballistic transition in a Persistent Random Walk}

\author{Amit Pradhan}
\cu
\author{Reshmi Roy}
\srm

\author{Purusattam Ray}
\imsc


\begin{abstract}

We study  persistent random walk with  time dependent velocity reversal probabilities and identify a criterion for a non-equilibrium dynamical transition. As a representative example, we consider a power law reversal probability $p(t)\sim t^{-\alpha}$ and show that the system undergoes a transition at $\alpha=1$, separating a super-diffusive regime for $\alpha<1$ from ballistic regime for $\alpha \geq 1$. Using the results for velocity correlations and persistence statistics, together with finite time scaling of the Binder cumulant and displacement fluctuations, we characterize the transition and its properties in detail. We further argue that the transition is not limited to the power law form, but can also arise for several other time dependent reversal probabilities satisfying the same criterion. The transition persists in arbitrary spatial dimensions provided isotropy of the velocity space is preserved.


\end{abstract}

\maketitle
Random walk models provide a fundamental framework for describing stochastic transport processes in many physical, chemical, and biological systems. In many such systems as well in natural systems, stochastic motion is influenced by memory effects and adaptive responses, leading to non stationary dynamics that cannot be captured by conventional Markovian random walk models. Among the various extensions of random walk dynamics, the persistent random walk (PRW) provides one of the simplest models incorporating temporal correlations in motion and has been extensively studied in the context of correlated stochastic processes \cite{renshaw,Hanneken,weiss,Kac}. Several variants of persistent random walks, including models with asymmetric transmission and correlated fluctuations, have been investigated in recent years \cite{rosetto,Sadjadi,Svenkeson,Marie}. Such correlated dynamics can strongly influence transport properties and may lead to transitions between different diffusion regimes \cite{andre}. Mathematical aspects of persistent random walks, including recurrence properties and scaling limits, have also been studied in detail, together with their connection to the telegrapher’s equation \cite{Cenac,cenac,Masoliver}.

Persistent random walk dynamics arises naturally in biological motion, particularly in the run and tumble behavior of bacteria such as E. coli, where organisms alternate between persistent motion and random reorientation events \cite{Berg,berg,Tailleur}. Similar ideas also appear in studies of search processes and exploration strategies in biological and ecological systems \cite{Benichou,benichou}. More broadly, persistent motion plays an important role in the physics of active matter, where self-propelled particles convert energy into directed motion. In this context, PRW type models are widely used to describe the dynamics of microswimmers, active colloids, and motile biological systems \cite{Marchetti,Bechinger,Elgeti,Solon}

Another important aspect of stochastic transport concerns anomalous diffusion, where the mean squared displacement grows nonlinearly in time. Such behavior has been extensively studied in Lévy walks, continuous time random walks, and related stochastic models with long-range correlations or heavy tailed statistics \cite{Zaburdaev,Metzler,metzler,Barkai}. In many systems these effects arise due to memory, heterogeneity, or non-stationary dynamics, leading to anomalous aggregation and aging phenomena \cite{Fedotov}. Aging effects, where statistical properties depend on the time elapsed since the system preparation, have been widely investigated in stochastic processes with long time memory \cite{barkai,Jeon,Schulz}.

For the conventional persistent random walk with constant flip probability $p$, the dynamics exhibits a temporal crossover from ballistic motion at short times to diffusive behavior at long times \cite{renshaw,Hanneken,weiss,Kac}. It is therefore natural to ask whether a dynamical transition can arise in persistent motion and, if so, what general condition controls such a transition. In particular, the long time dynamics can change qualitatively depending on whether the total number of velocity reversals grows with time without bound or remains finite. Understanding this mechanism and its consequences for transport is the central motivation of the present work.

In this work we study a persistent random walk with a time dependent velocity reversal probability $p(t)$, leading to explicitly nonstationary and aging dynamics.
Specifically, we consider a one dimensional discrete time process in which the velocity $v(t)=\pm1$ evolves according to 
\begin{equation}
v(t+1)=
\begin{cases}
\;\;v(t), & \text{with probability } 1-p(t+1),\\[4pt]
-\,v(t), & \text{with probability } p(t+1),
\end{cases}
\end{equation}
and the position is updated as 
\begin{equation}
    x(t+1) = x(t) + v(t+1).
 \label{position update}   
\end{equation}


For an arbitrary time dependent flip probability $p(t)$, Defining the  velocity persistence parameter $\gamma(t)=1-2p(t)$, the dynamics obeys exact recursion relations
\begin{equation}
\label{recurrence for A(t)}
    A(t+1) = \gamma(t+1)A(t) +1 \quad ; \quad A(0) = 0,
\end{equation}
and
\begin{equation}
\label{recurrence for MSD}
    \langle x^2(t+1)\rangle = \langle x^2(t)\rangle + 2\gamma(t+1) A(t) + 1,\quad \langle x^2(0)\rangle = 0,
\end{equation}
where $A(t) = \langle x(t)v(t)\rangle$. Iteration of these relations yields the closed form expressions
\begin{equation}
\label{A(t)_MSD}
    A(t) = \sum_{s=1}^{t}\prod_{u=s+1}^{t}\gamma(u), \quad \langle x^2(t)\rangle = t + 2 \sum_{s=0}^{t-1} \gamma(s+1) A(s), 
\end{equation}
with the empty product convention 
$\prod_{u=t+1}^{t}\gamma(u) = 1$.
The same persistence product determines the two time velocity autocorrelation,
\begin{equation}
\label{velocity autocorrelation}
    \langle v(t)v(t^\prime)\rangle = \prod_{u=t^\prime+1}^{t} \gamma(u), \quad t>t^\prime,
\end{equation}
implying nonstationary, aging dynamics whenever $p(t)$ decays with time.

The long time behavior is controlled by the expected number of velocity reversals up to time $t$,
\begin{equation}
\mathcal{N}(t) = \sum_{u=1}^{t}p(u),
\end{equation}
which determines whether the walker continues to change direction indefinitely or eventually settles into a fixed direction. Using Eq. (\ref{A(t)_MSD}), one finds that when $N(\infty) = \sum_{u=1}^{\infty} p(u)$ remains finite, the velocity correlations remains finite at long times, leading  to ballistic motion,
\begin{equation}
\label{universal ballistic behavior}
    \langle x^2(t)\rangle \sim t^2,
\end{equation}
independent of the detailed form of the reversal probability.

On the other hand, when $\mathcal{N}(t)$ grows without bound as $t\to \infty$, and the reversal probability decays slower than $1/t$, such that $tp(t)\gg 1$, the dominant contribution to the sum for $A(t)$ in Eq. (\ref{A(t)_MSD}) comes from values of $s$ close to $t$. In this regime,
\begin{equation}
    A(t) \sim \frac{1}{p(t)}, \quad t\to \infty.
\end{equation}
Consequently, the mean squared displacement asymptotically scales as 
\begin{equation}
\label{General relation for msd}
    \langle x^2(t)\rangle \sim \sum_{u\gg1}^{t} \frac{1}{p(u)}.
\end{equation}
Thus a dynamical transition can arise for reversal protocol in which $\mathcal{N}(\infty)$ remains finite in one regime but diverges in another.


As a representative example of such reversal protocols, we consider a velocity reversal probability of the power law form
\begin{equation}
    p(t) = \frac{c}{t^{\alpha}},
\end{equation}
and show that it produces a dynamical transition at $\alpha=1$, separating a super-diffusive regime for $\alpha<1$ from ballistic motion for $\alpha\geq 1$. Below we discuss the different asymptotic regimes for the power law case by varying $\alpha$.


\textbf{Case I ($\boldsymbol{\alpha<1}$):}
    For $\alpha<1$, the velocity autocorrelation [Eq. (\ref{velocity autocorrelation})] decays rapidly at long times. For $t\to \infty$ and $t^\prime\gg1$ with $t^\prime<t$,
    \begin{equation}
        \langle v(t)v(t^\prime)\rangle \sim \exp\left[-\frac{2c}{1-\alpha}(t^{1-\alpha}-{t^\prime}^{1-\alpha})\right],
    \end{equation}
    indicating a stretched exponential decay of correlation. Consequently velocity reversals persist indefinitely, and the expected number of flips up to time $t$ grows as
    \begin{equation}
        \mathcal{N}(t) \sim \sum_{u=1}^{t} u^{-\alpha} \sim t^{1-\alpha}.
    \end{equation}
    Since $\mathcal{N}(t)$ diverges for $\alpha<1$ and $tp(t) \sim t^{1-\alpha}\gg 1$, Eq. (\ref{General relation for msd}) gives
    \begin{equation}
    \label{MSD_alpha<1}
        \langle x^2(t)\rangle \xrightarrow[t\to \infty]{} \sum_{u\gg 1}^{t}u^{\alpha}\sim \frac{1}{c(\alpha+1)}t^{\alpha+1},
    \end{equation}
    implying super-diffusive motion [Inset (a) of Fig. \ref{MSD vs t for alpha = 0.5,1,1.5}].
    
     The survival probability for a run of length $l$
    \begin{equation}
        P(l) = \prod_{t=1}^{l} \left(1-\frac{c}{t^{\alpha}}\right),
    \end{equation}
    has the asymptotic form
    \begin{equation}
    \label{P(l)_alpha<1}
 P(l) \sim \exp\left[-\frac{c}{1-\alpha}l^{1-\alpha}\right], . 
\end{equation}
leading to a persistence length distribution (inset (a) of Fig. \ref{path length dist for alpha = 0.6,1,1.5} in \cite{SM}) 
\begin{equation}
\label{tail of F(l)}
 F(l) = P(l-1)\left(\frac{c}{l^{\alpha}}\right) \sim \left(\frac{c}{l^{\alpha}}\right)  \exp\left[-\frac{c}{1-\alpha}l^{1-\alpha}\right].  
\end{equation}
The stretched exponential cutoff ensures a finite but increasing mean run length, so that longer runs become more prominent as $\alpha$ increases, although they are ultimately interrupted by the diverging number of reversals.

\begin{figure}[h]
    \centering
    \includegraphics[width=\linewidth]{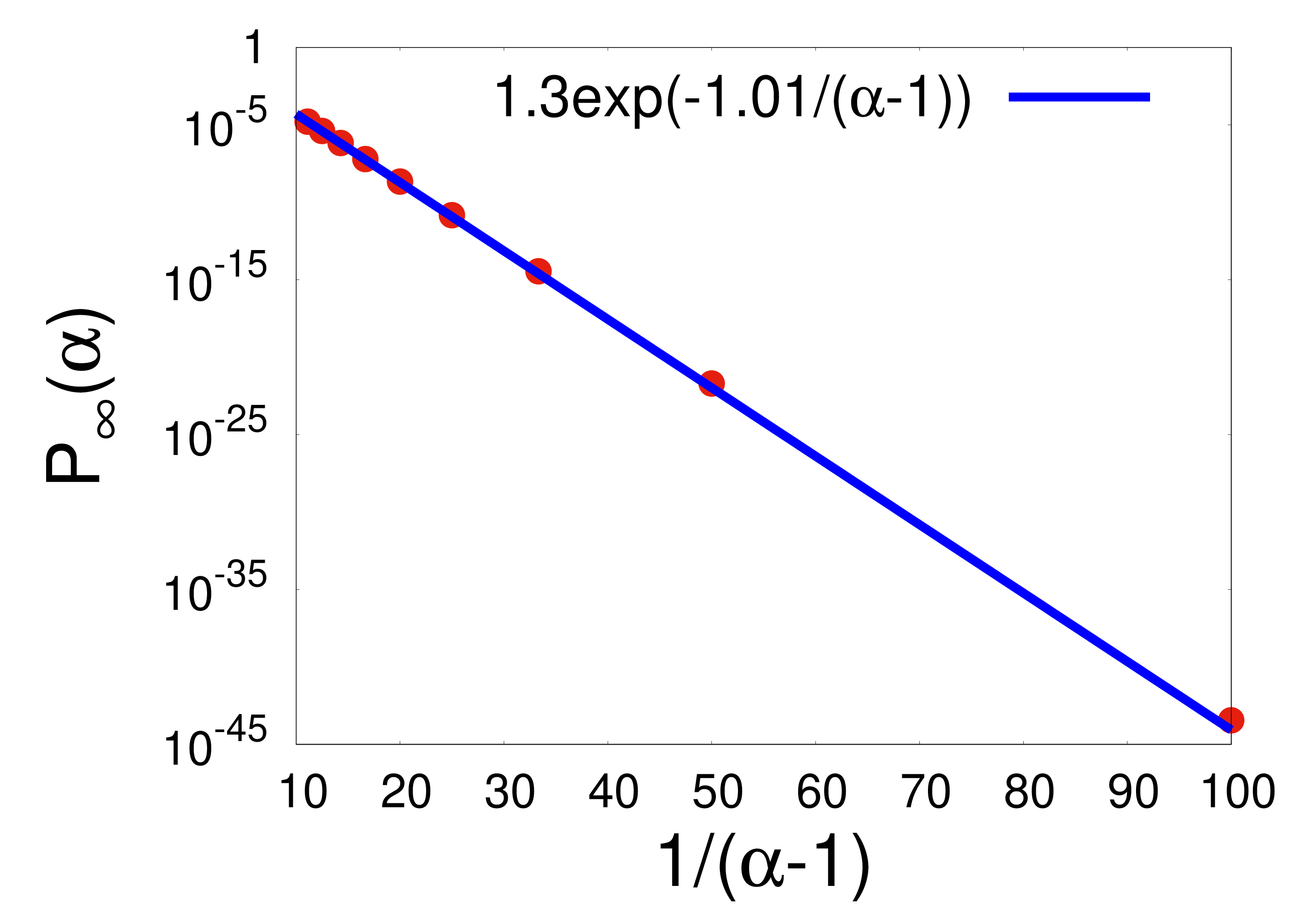}
    \caption{Plot of the survival probability $P_{\infty}(\alpha)$ as a function of $1/(\alpha-1)$ in log-linear scale for $\alpha>1$. The numerical data are well described by the analytical form $P_{\infty}(\alpha)\sim \exp[-1/(\alpha-1)]$ [Eq. (\ref{order parameter}) with $c=1$ and $\zeta(\alpha)\sim 1/(\alpha-1)$ near $\alpha=1^+$], indicating that $P_{\infty}$ decays continuously and exponentially as $\alpha\to1+$. The solid red line shows the best fit $P_{\infty}(\alpha)\approx 1.3\exp[-1.01/(\alpha-1)]$, in excellent agreement with the analytical prediction.}
    \label{survival_vs_alpha}
\end{figure}

    \textbf{Case II ($\boldsymbol{\alpha>1}$):}
    For $\alpha>1$, in the limit $t\to \infty$ with $t^\prime\gg1$ fixed, the velocity autocorrelation approaches a finite value,
     \begin{equation}
    \langle v(t)v(t^\prime)\rangle \quad \rightarrow \quad  \exp\left[-2c\sum_{u=t^\prime+1}^{\infty}u^{-\alpha}\right].
   \end{equation}
    indicating persistent memory of the initial velocity. In this regime, the total number of velocity reversals converges,
    \begin{equation}
        \mathcal{N}(\infty) \sim \sum_{u=1}^{\infty}u^{-\alpha} <\infty,
    \end{equation}
    so that with finite probability the velocity never flips after some random time. Since $\mathcal{N}(\infty)$ remains finite for $\alpha>1$, Eq. (\ref{universal ballistic behavior}) immediately implies ballistic growth of the mean squared displacement (inset (b) of Fig. \ref{MSD vs t for alpha = 0.5,1,1.5}),
    \begin{equation}
    \label{MSD for alpha>1}
    \langle x^2(t)\rangle \xrightarrow[t\to\infty]{} C_0t^2, \quad C_0 = e^{-2c\zeta(\alpha)}.
    \end{equation}
     The survival probability saturates to a constant,
    \begin{equation}
    \label{order parameter}
        P(l)\xrightarrow[l\to\infty]{} P_{\infty}(\alpha) \sim \exp[-c\zeta(\alpha)],
    \end{equation}
    Here $\zeta(\alpha)$ denotes the Riemann zeta function which diverges at $\alpha= 1$ as $\zeta(\alpha)\sim 1/(\alpha-1)$. Unlike the case $\alpha<1$, where $P(l)$ vanishes as $l\to \infty$ [Eq. \ref{P(l)_alpha<1}], here a finite fraction $P_\infty(\alpha)$ of trajectories never flips after some time and moves ballistically with $|x(t)|\sim t$, while the remaining trajectories undergo infinitely many reversals and contribute sub-ballistically. Consequently,
   \begin{equation}
       \lim_{t\to \infty}\frac{\langle |x|\rangle}{t} = P_\infty(\alpha),
   \end{equation}
   showing that $P_\infty(\alpha)$  acts as a ballistic order parameter [Fig. \ref{survival_vs_alpha}].
    
    The persistence length distribution behaves as (inset (b) of Fig. ~\ref{path length dist for alpha = 0.6,1,1.5} in \cite{SM}]
    \begin{equation}
    F(l) = P(l-1)\left(\frac{c}{l^{\alpha}}\right) \sim cP_{\infty}(\alpha) l^{-\alpha}.
   \end{equation}

    \textbf{Case III ($\boldsymbol{\alpha=1}$):}
    The marginal case $\alpha=1$ separates these two regimes. Here the velocity autocorrelation decays algebraically
    \begin{equation}
    \label{velocity autocorrelation at alpha = 1}
        \langle v(t)v(t^\prime)\rangle \sim \left(\frac{t^\prime}{t}\right)^{2c}, \quad t\to \infty,\quad t^\prime>>1,
    \end{equation}
    indicating a slow, power-law decay of correlation in time. In this case
    \begin{equation}
        \mathcal{N}(t) \sim \sum_{u=1}^{t} \frac{1}{u} \sim \ln t,
    \end{equation}
    so that the expected number of velocity reversals diverges logarithmically as $t\to \infty$. Since $p(t)\sim 1/t$, the condition $tp(t)\gg 1$ is no longer satisfied and therefore Eqs. (\ref{universal ballistic behavior}) and (\ref{General relation for msd}) can no longer be applied. However, a detailed calculation for $\alpha=1$ shows that the slow decay of temporal correlations remains sufficient to sustain ballistic transport (Fig. \ref{MSD vs t for alpha = 0.5,1,1.5}), giving
     \begin{equation}
    \label{MSD_alpha=1}
        \langle x^2(t)\rangle \sim \frac{1}{2c+1}t^2.
    \end{equation}
    The survival probability and persistence length  distribution behave as (Fig. \ref{MSD vs t for exponential and linear flip} in \cite{SM}),
    \begin{equation}
        P(l) \sim l^{-c}, \quad F(l) \sim l^{-(1+c)}.
    \end{equation}

    Thus, the transition at $\alpha=1$ is driven by a qualitative change in the temporal decay of velocity correlations : from rapidly decaying ($\alpha<1$), to marginal ($\alpha=1$), to persistent ($\alpha>1$). While ballistic motion for $\alpha>1$ arises from velocity freezing, at $\alpha=1$ it is generated by long-lived temporal correlations, making the two regimes dynamically distinct.


\begin{figure}[h]
    \includegraphics[width=\linewidth]{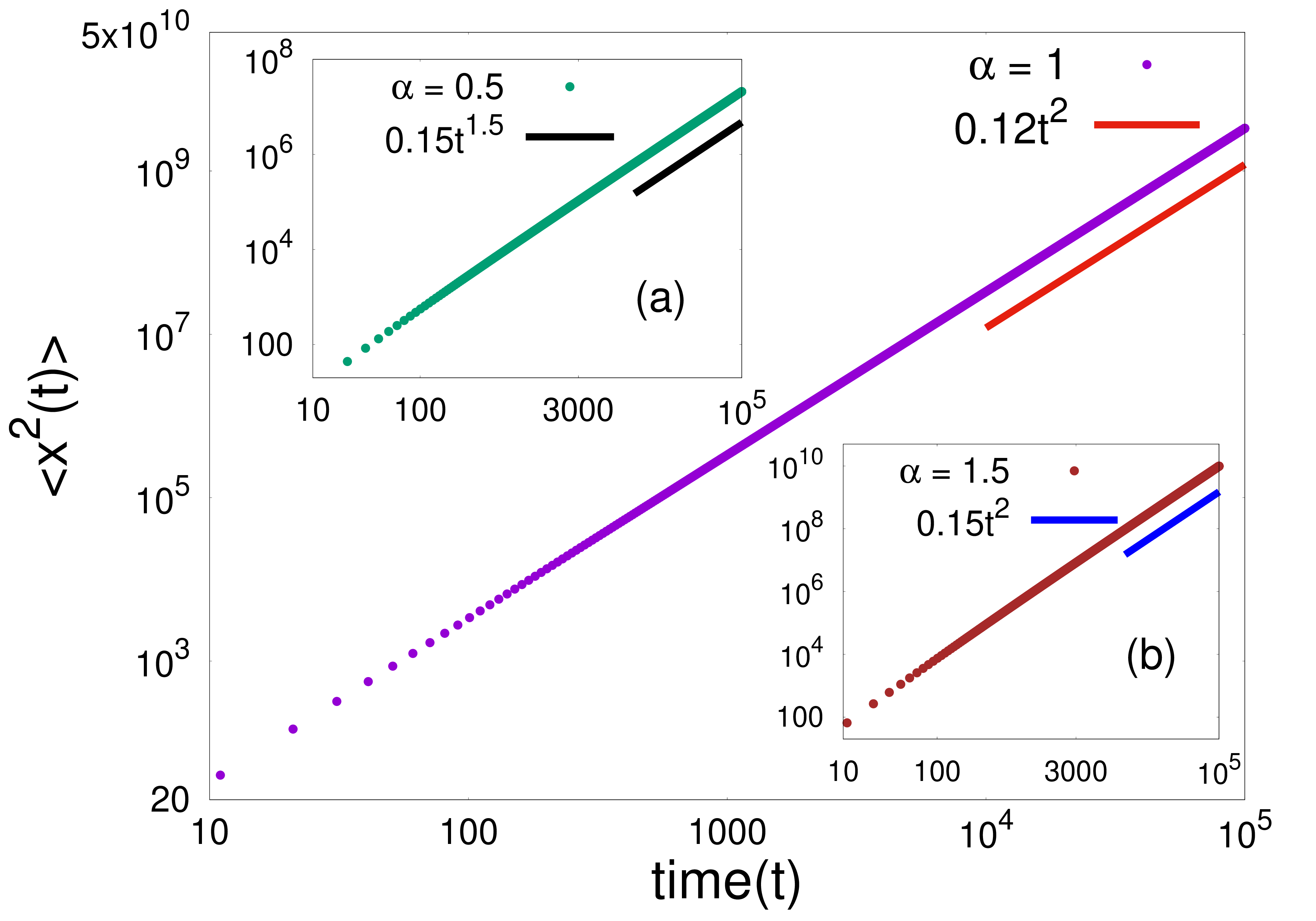}
    \caption{Mean squared displacement $\langle x^2(t)\rangle$ for a persistent random walk with power law flip probability $p(t)=c/t^{\alpha}$ shown on a log-log scale. Data points correspond to numerical simulations, while solid lines indicate analytical predictions [Eqs. (\ref{MSD_alpha<1}), (\ref{MSD_alpha=1}), and (\ref{MSD for alpha>1}) ]. The main panel shows the marginal case $\alpha=1$, displaying ballistic scaling $\langle x^2(t)\rangle \sim t^2$. Inset (a) corresponds to $\alpha = 0.5$, exhibiting super-diffusive behavior $\langle x^2(t)\rangle \sim t^{1+\alpha}$, while the inset (b) shows $\alpha=1.5$, for which ballistic growth persists with a non universal prefactor.}
    \label{MSD vs t for alpha = 0.5,1,1.5}
\end{figure}

This transition is directly characterized by the growth of displacement fluctuations and by the crossing of the Binder cumulant curves at the critical point $\alpha_c=1$. To quantify the displacement fluctuations, we define the variance of the displacement, $\sigma_x^2(t)=\langle x^2(t)\rangle-\langle x(t)\rangle^2$, which, for an arbitrary time dependent flip probability $p(t)$, can be written exactly as
\begin{equation}
    \sigma_x^2(t) = t+ 2\sum_{u=1}^{t}\sum_{r=1}^{u-1}\prod_{k=u-r+1}^{u}\gamma(k), \quad \gamma(k) = 1-2p(k).
\end{equation}

\begin{figure}[h!]
    \centering
    \includegraphics[width=\linewidth]{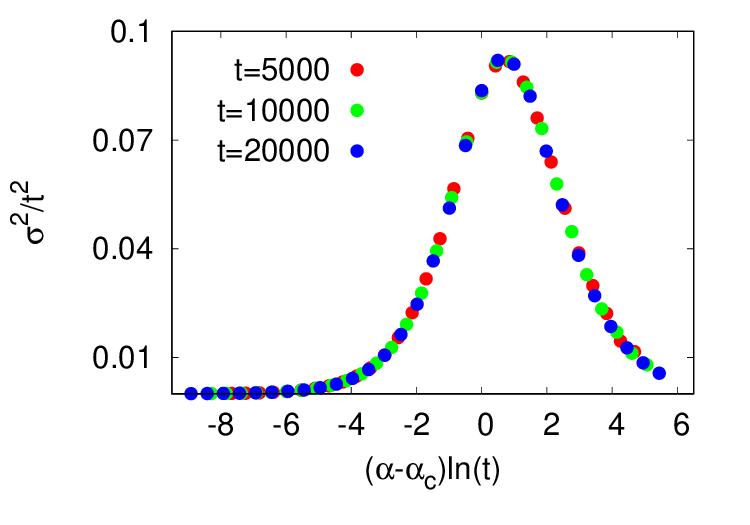}
    \caption{Finite time scaling of the displacement variance $\sigma_x^2 = \langle x^2\rangle-\langle x \rangle^2$ computed over one half of the position distribution. The data for $3$ different times at $t=5000,10000,20000$ collapse onto a single universal curve when plotted according to the proposed scaling form in Eq. (\ref{Finite time scaling of variance}). }
    \label{varience_collapse}
\end{figure}

For the power law form $p(t)=ct^{-\alpha}$, the variance admits an asymptotic scaling form near the transition (Sec. SVI in \cite{SM}),
\begin{equation}
\label{Finite time scaling of variance}
    \sigma_x^2(t) \sim t^2 \mathcal{G}[|\alpha-1|\ln t],
\end{equation}
where the scaling function $\mathcal{G}$, is defined by the following integral
\begin{equation}
\label{scaling function}
    \mathcal{G}[(\alpha-1)\ln t] =\int_{0}^{1}\frac{\tau^{2c}e^{-[c(\alpha-1)(\ln \tau)^2]}e^{-[2c(\alpha-1)\ln t \ln \tau]}}{1-c(\alpha-1)\ln \tau}d\tau.
\end{equation}
This establishes that the fluctuations obey a logarithmic finite-time scaling, rather than a conventional power-law form. Consistently, numerical data for different times collapse onto a universal curve when plotted against $(\alpha-1)\ln t$ (Fig. \ref{varience_collapse}), in excellent agreement with the analytical prediction. In addition, the displacement distribution at criticality becomes exceptionally broad (see Fig. ~\ref{path length dist for alpha = 0.95} in \cite{SM}), directly reflecting the strong fluctuations associated with the transition.

The logarithmic scaling variable implies an exponentially large characteristic timescale 
\begin{equation}
    t^*(\alpha) \sim e^{\frac{1}{|\alpha-1|}},
\end{equation}
which characterizes the observation time up to which the dynamics remains effectively indistinguishable from the marginal case $\alpha=1$. For $\alpha>1$, this scale has a direct physical interpretation : the fraction $P_{\infty}(\alpha)$ of permanently ballistic trajectories  vanishes exponentially near the transition, $P_\infty(\alpha)\sim e^{-1/(\alpha-1)}$ [Eq. (\ref{order parameter}) at $\alpha=1^+$], implying $t^*(\alpha)=1/P_\infty(\alpha)$. The divergence of $t^*$ at $\alpha=1$ reflects the  absence of such trajectories at criticality.

\begin{figure}[h]
    \centering
    \includegraphics[width=\linewidth]{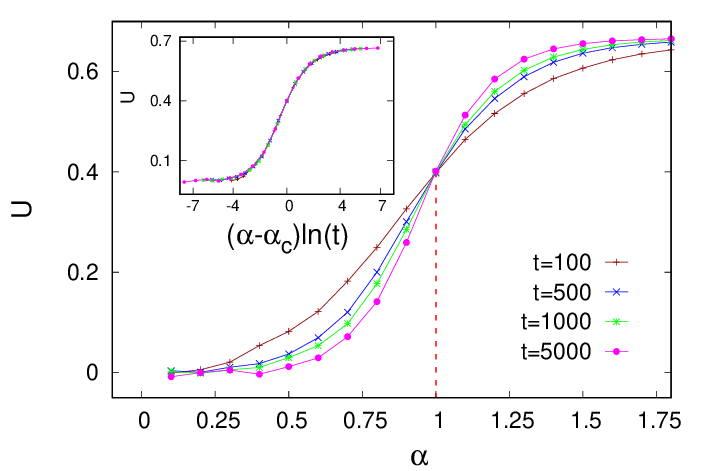}
    \caption{The Binder cumulant $U$ is shown against $\alpha$ for different times. The inset shows the data collapse of the binder cumulant for different times when $\alpha$ is scaled as $(\alpha-\alpha_c)\ln t$ with $\alpha_c=1$. The red dashed line indicates the transition at $\alpha_c=1$ where the order of the curves change. These data are for time $t=100, 500, 1000$ and  $5000$ as mentioned in the legends of the main plot, averaged over $10^5$ different realisations.}
    \label{binder_cumulant}
\end{figure}

 The Binder cumulant of the displacement,
\begin{equation}
    U(\alpha,t) = 1-\frac{\langle x^4\rangle}{3\langle x^2\rangle^2},
\end{equation}
exhibits a sharp crossing of curves for different observation times at $\alpha_c=1$ (Fig. \ref{binder_cumulant}) and collapses under the same logarithmic scaling variable. This provides clear evidence that the change at $\alpha=1$ is a genuine dynamical phase transition. 



We now generalize the persistent random walk with time dependent flip probability to $d$ spatial dimensions, assuming isotropy of the velocity space. The velocity is a discrete vector taking values on the $d$ dimensional hypercubic lattice,
\begin{equation}
    \mathbf{v}(t) \in \{\pm \mathbf{\hat{e_1}},\mathbf{\hat{\pm e_2}},....,\mathbf{\hat{\pm e_d}}\},
\end{equation}
where $\mathbf{\hat{e_i}}$ are the Cartesian unit vectors, so that $|\mathbf{v}(t)|^2=1$ at all times. Given the velocity $\mathbf{v}(t)$ at time $t$, the update rule is defined as follows : with probability $1-p(t+1)$ the velocity persists, $\mathbf{v}(t+1)=\mathbf{v}(t)$, while with probability $p(t+1)$ the particle tumbles and chooses a new direction uniformly from the remaining $2d-1$ directions, ensuring the isotropy of the dynamics.

With this isotropic update rule, the structure of the recursion relations for the position velocity correlation $A(t) = \langle \mathbf{x}(t)\cdot \mathbf{v}(t)\rangle$ [Eq. (\ref{recurrence for A(t)})] and the mean squared displacement $r^2(t) = |\mathbf{x}(t)|^2$ [Eq. (\ref{recurrence for MSD})] where the position vector is
\begin{equation}
    \mathbf{x}(t) = \{x_1(t),x_2(t).....x_d(t)\}, \quad \text{with} \quad x_i(t) \in [-t,t],
\end{equation}
is identical to the one dimensional case. The only modification is the replacement of the velocity persistence parameter 
\begin{equation}
    \gamma(t) = 1-2p(t) \quad \to \quad  \gamma_d(t)=1-\frac{2d}{2d-1}p(t).
\end{equation}
Consequently, all asymptotic results derived in one dimension remain valid in arbitrary dimension under this replacement. In particular, the exact forms of Eqs. (\ref{recurrence for A(t)})-(\ref{A(t)}) are unchanged, implying that the super-diffusive to ballistic transition at $\alpha=1$ is dimension independent, with only nonuniversal prefactors depending on $d$. This conclusion holds provided rotational symmetry of the velocity space is preserved.

Finally, we contrast the power law flip probability with other commonly studied cases. 
For the conventional persistent random walk with constant flip probability $p$, the dynamics exhibits a temporal crossover from ballistic motion at short times to diffusive behavior at long times, as established in earlier studies \cite{renshaw,Hanneken,weiss,Kac}. For exponentially decaying flips, $p(t)=p_0e^{-\lambda t}$, a finite time scale $\tau\sim \lambda^{-1}$ exists beyond which velocity reversals are exponentially suppressed (Sec. ~\ref{MSD_scaling_expo} in \cite{SM}), while for linearly decreasing flips, $p(t)=p_0-\beta t$, reversals cease after a finite cutoff time $\tau=p_0/\beta$ (Sec. ~\ref{MSD_scaling_linear} in \cite{SM}). As a result, the expected number of velocity reversals $\mathcal{N}(\infty)$ remains finite due to the presence of the characteristic timescale, leading to ballistic motion at long times together with a crossover from short-time diffusive to long-time ballistic behavior, rather than a sharp dynamical transition.

For reversal protocols which do not have any intrinsic time scale, other than power law case, there can be several forms which still satisfy the condition for the transition and lead to a sharp transition, such as $p(t)=c \log t/t^{\alpha},p(t) = c/[t^{\alpha}\log t],p(t) = c/[t(\log t)^{\alpha}],p(t)=c(\log t)^{\alpha}/t,$ and $p(t)=c/[t \log t(\log \log t)^{\alpha}].$ For all the above forms except $p(t)=c(\log t)^{\alpha}/t$, the expected number of velocity reversals up to time $t$, $\mathcal{N}(t)$ grows without bound and diverges as $t\to \infty$ for $\alpha\leq 1$, and remains finite for $\alpha>1$, leading to a transition from diffusive to ballistic motion at $\alpha=1$. For $p(t) = c(\log t)^{\alpha}/t$, $\mathcal{N}(t)$ remains finite for $\alpha<-1$ and grows without bound for $\alpha\geq -1$, giving a transition at $\alpha=-1$. These results show that the transition is not limited to the simple power law form, but arises generally whenever the expected number of velocity reversals grows with time without bound in one regime and remains finite in another.


In summary, we have studied persistent random walks with time dependent velocity reversal probabilities, and identified a general criterion for a sharp dynamical transition in such systems. As a representative example, we considered a power law reversal probability and showed that it leads to a transition at $\alpha=1$, separating super-diffusive and ballistic regimes. We further showed that similar transitions can also arise for other reversal protocols
whenever the expected number of velocity reversals grows with time without bound in one regime but remains finite in another.

For the power law case, we have characterized the transition in detail. Ballistic motion arises through two distinct mechanisms: velocity freezing for $\alpha>1$ and long-lived velocity correlations at the critical point $\alpha=1$.


The transition persists in arbitrary spatial dimensions, provided isotropy of the velocity space is maintained, demonstrating the robustness of the phenomenon.

\textit{Acknowledgements}---
AP would like to acknowledge University Grant Commission (UGC), Govt. of India for financial support (Student ID: 241610061476). RR acknowledges SRM University-AP for the University Post doctoral fellowship. AP and RR are also grateful to Arghya Das for useful discussions.

\newpage 
\cleardoublepage

\onecolumngrid
\begin{center}
\textbf{\large Supplemental Material for\\
``Diffusive-to-Ballistic transition in a Persistent Random Walk''}\\[5pt]

\begin{center}
 {\small Amit Pradhan$^{1}$, Reshmi Roy$^{2}$ and Purusattam Ray$^{3}$}  
\end{center}

\begin{center}
{\sl \footnotesize
$^{1}$Department of Physics, University of Calcutta, 92 Acharya Prafulla Chandra Road, Kolkata 700009

$^{2}$Department of Physics, SRM University - AP, Amaravati, Andhra Pradesh - 522240, India

$^{3}$Institute of Mathematical Sciences, CIT Campus, Chennai 600113, India}
\end{center}
\end{center}

\bigskip
\begin{quote}
This Supplemental Material provides detailed derivations of some of our main results of the paper. Moreover, it also provides additional discussions that support our finding as announced in the main text. 
\end{quote}

\vspace*{0.3cm}

\setcounter{equation}{0}
\renewcommand{\theequation}{S\arabic{equation}}
\setcounter{figure}{0}
\renewcommand{\thefigure}{S\arabic{figure}}
\setcounter{table}{0}
\setcounter{page}{1}
\setcounter{section}{0}
\setcounter{secnumdepth}{2}
\renewcommand{\thesection}{S\Roman{section}}
\renewcommand{\thepage}{\arabic{page}}
\makeatletter

\section{Derivation of the exact relations for position velocity correlations $\boldsymbol{A(t)}$ and mean square displacement $\boldsymbol{\langle x^2(t)\rangle}$}

In this section we derive the exact relations for the position velocity correlations $A(t)=\langle x(t) v(t)\rangle$ and the mean square displacement $\langle x^2(t)\rangle$ used in the main text in Eq. (5). These relations follow directly from the microscopic update rule of the persistent random walk with time dependent velocity reversal probability $p(t)$

Starting from the update rule of the walker position
\begin{equation}
    x(t+1) = x(t) + v(t+1),
\end{equation}
where $v(t)=\pm 1$, the velocity flips with probability $p(t+1)$ and remain unchanged with probability $1-p(t+1)$. Taking the average over this stochastic update and over realizations gives the recurrence relation
\begin{equation}
    \langle v(t+1)\rangle = (1-2p(t+1))\langle v(t)\rangle.
\end{equation}
Defining $\gamma(t)=1-2p(t)$, this can be written as
\begin{equation}
\label{mean velocity recurrence}
    \langle v(t+1)\rangle = \gamma(t+1) \langle v(t)\rangle .
\end{equation}
Taking the expectation value of the position update rule yields
\begin{equation}
    \langle x(t+1)\rangle = \langle x(t)\rangle + \langle v(t+1)\rangle.
\end{equation}
Multiplying the position update equation by $v(t+1)$ and averaging gives
\begin{equation}
    A(t+1) = \langle x(t)v(t+1)\rangle + 1,
\end{equation}
where $A(t)=\langle x(t)v(t)\rangle$. Using the conditional relation $\langle x(t)v(t+1)\rangle = \gamma(t+1)\langle x(t)v(t)\rangle$, we obtain the exact recurrence
\begin{equation}
 A(t+1) = \gamma(t+1)A(t) + 1, \quad A(0) = 0,   
\end{equation}
which corresponds to Eq. (3) of the main text. Iterating this relation yields
\begin{equation}
    A(t) = \sum_{s=1}^{t}\prod_{u=s+1}^{t}\gamma(u),
    \label{A(t)}
\end{equation}
with the empty product convention $\prod_{u=t+1}^{t}\gamma(u) = 1$.

The mean square displacement follows from squaring the position update equation,
\begin{equation}
  x^2(t+1) = x^2(t) + 2x(t)v(t+1) + v^2(t+1).  
\end{equation}
Taking the expectation value and using $v^2(t)=1$ gives
\begin{equation}
    \langle x^2(t+1)\rangle = \langle x^2(t)\rangle + 2\langle x(t)v(t+1)\rangle + 1.
\end{equation}
Using again $\langle x(t)v(t+1)\rangle = \gamma(t+1)\langle x(t)v(t)\rangle$, we obtain
\begin{equation}
    \langle x^2(t+1)\rangle = \langle x^2(t)\rangle + 2\gamma(t+1)A(t)+1,
\end{equation}
which is Eq. (4) of the main text. Summing this relation from $t=0$ to $t-1$ and using $\langle x^2(0)\rangle=0$ gives
\begin{equation}
\label{MSD}
    \langle x^2(t)\rangle = t + 2\sum_{s=0}^{t-1}\gamma(s+1)A(s),
\end{equation}
which corresponds to Eq. (5) of the main text.

\section{General asymptotic scaling in terms of the expected number of velocity reversals}

From Eq. (\ref{A(t)}), we define
\begin{equation}
    Q(s,t) = \prod_{u=s+1}^{t}(1-2p(u)),
\end{equation}
so that
\begin{equation}
\label{A(t)_in_terms_of_Q}
    A(t) = \sum_{s=1}^{t} Q(s,t).
\end{equation}
In the asymptotic limit $t\to \infty$, $p(t)\ll 1$. Using $\ln(1-2p(u)) \approx -2p(u)$, one obtains
\begin{equation}
    Q(s,t) \sim \exp\left[-2\sum_{u=s+1}^{t}p(u)\right].
\end{equation}
Defining the expected number of velocity reversals up to time $t$,
\begin{equation}
    \mathcal{N}(t) = \sum_{u=1}^{t}p(u),
\end{equation}
$Q(s,t)$ takes the asymptotic form
\begin{equation}
    Q(s,t) \sim \exp[-2(\mathcal{N}(t)-\mathcal{N}(s))].
\end{equation}
The asymptotic behavior is therefore controlled by the large $t$ behavior of $\mathcal{N}(t)$.

\vspace{5mm}

\textbf{Case A ($\boldsymbol{\mathcal{N}(\infty)<\infty}$):}

\vspace{5mm}

We first consider the case 
\begin{equation}
    \mathcal{N}(\infty) = \sum_{u=1}^{\infty}p(u)<\infty.
\end{equation}
To analyze the asymptotic behavior of $A(t)$, we split the sum as 
\begin{equation}
    A(t) = \sum_{s=1}^{S}Q(s,t) + \sum_{s=S+1}^{t}Q(s,t),
\end{equation}
where $S\gg1$ is fixed. The first contribution remains finite as $t\to \infty$. For the second contribution, both $s$ and $t$ are large, and since $\mathcal{N}(\infty)$ converges, $\mathcal{N}(t)-\mathcal{N}(s) \to 0,$, implying $Q(s,t)\to 1$. Therefore,
\begin{equation}
    A(t) \sim \sum_{s=S+1}^{t} 1 \sim t.
\end{equation}
Using the exact relation of MSD given in Eq. (\ref{MSD}), together with $A(t)\sim t$, one immediately obtains 
\begin{equation}
    \langle x^2(t)\rangle \sim t^2.
\end{equation}
Thus, whenever $\mathcal{N}(\infty)$ remains finite, the asymptotic motion is ballistic independent of the detailed form of $p(t)$.

\vspace{5mm}

\textbf{Case B ($\boldsymbol{\mathcal{N}(\infty)=\infty}$):}

 \vspace{5mm}

 We now consider the case where 
 \begin{equation}
     \mathcal{N}(t) \to \infty \quad \text{as} \quad  t \to \infty.
 \end{equation}
In this regime, the dominant contribution to the sum in Eq. (\ref{A(t)_in_terms_of_Q}) comes from $s$ close to $t$. writing $s=t-r$, with $r\ll t$, we Taylor expand $\mathcal{N}(s)$ about $s=t$,
\begin{equation}
    \mathcal{N}(t-r) = \mathcal{N}(t) -rN^{\prime}(t) + \mathcal{O}(r^2).
\end{equation}
Since $\mathcal{N}(t) = \sum_{u=1}^{t} p(u)$, its continuum derivative satisfies asymptotically $\mathcal{N}^{\prime}(t) \approx p(t)$. Keeping only the leading term,
\begin{equation}
    \mathcal{N}(t) - \mathcal{N}(t-r) \approx rp(t),
\end{equation}
which gives
\begin{equation}
    Q(t-r,t) \sim e^{-2rp(t)}.
\end{equation}
The exponential suppression identifies the characteristic scale $r_c\sim 1/p(t)$, beyond which the contribution to the sum becomes negligible. Since, the Taylor expansion requires $r\ll t$, the asymptotic scaling holds provided
\begin{equation}
    \frac{1}{p(t)} \ll t \quad  \Rightarrow \quad tp(t)\gg 1.
\end{equation}
Thus, asymptotically for $t\to \infty$,
\begin{equation}
    A(t) \sim \sum_{r=0}^{1/p(t)} e^{-2rp(t)}.
\end{equation}
For $p(t)\ll 1$, the geometric sum gives
\begin{equation}
    A(t) \sim \frac{1}{p(t)}.
\end{equation}
Substituting this into the exact relation of MSD given in Eq. (\ref{MSD}), one obtains asymptotically
\begin{equation}
\label{General MSD scaling}
    \langle x^2(t)\rangle \sim \sum_{u\gg1}^{t} \frac{1}{p(u)}.
\end{equation}
Thus, whenever $\mathcal{N}(t)$ grows with time without bound and the additional condition $tp(t)\gg 1$ is satisfied, the asymptotic scaling of the MSD is determined by Eq. (\ref{General MSD scaling}).

\section{Velocity Autocorrelation for Power Law Reversal Dynamics}

In this section, we derive the two time velocity autocorrelation function for a persistent random walk with a time dependent velocity reversal probability $p(t)=c/t^{\alpha}$. Starting from the recurrence relation for the mean velocity, we obtain an exact product form for the autocorrelation and analyze its asymptotic behavior in different regimes of $\alpha$.

In Eq. (\ref{mean velocity recurrence}), the recurrence relation for the mean velocity is given. The two time velocity autocorrelation, $\langle v(t)v(t^\prime)\rangle$, satisfies an analogous relation
\begin{equation}
    \langle v(t+1)v(t^\prime)\rangle = \gamma(t+1) \langle v(t)v(t^\prime)\rangle,
\end{equation}
which upon iteration from time $t^\prime$ to $t>t^\prime$, yields the product form
\begin{equation}
    \langle v(t)v(t^\prime)\rangle = \prod_{u=t^\prime+1}^{t}\gamma(u),
\end{equation}
which corresponds to Eq. (6) of the main text. For $p(u)=c/u^{\alpha}$, this becomes
\begin{equation}
  \langle v(t)v(t^\prime)\rangle = \prod_{u=t^\prime+1}^{t} \left(1-\frac{2c}{u^\alpha}\right).
\end{equation}
For $t\to \infty$ and $t^\prime \gg1$, we expand the logarithm to leading order in $u^{-\alpha}$
\begin{equation}
    \ln \langle v(t)v(t^\prime)\rangle = \sum_{u=t^\prime+1}^{t}\ln\left(1-\frac{2c}{u^{\alpha}}\right) \approx -2c\sum_{u=t^\prime+1}^{t}u^{-\alpha}.
\end{equation}
This yields the asymptotic form
\begin{equation}
    \langle v(t)v(t^\prime)\rangle \sim \exp\left[-2c\sum_{u=t^\prime+1}^{t}u^{-\alpha}\right].
\end{equation}
The behavior of the autocorrelation is governed by the asymptotics of the sum $\sum_{u=t^\prime+1}^{t}u^{-\alpha}$, leading to three distinct regimes:
\begin{description}
    \item[(i) $\boldsymbol{\alpha<1}$ : Stretched Exponential Decay] 
    For $\alpha<1$,
    \begin{equation}
        \sum_{u=t^\prime+1}^{t}u^{-\alpha} \sim \frac{1}{1-\alpha}\left[t^{1-\alpha}-(t^\prime)^{1-\alpha}\right].
    \end{equation}
    Thus, for $t\to \infty$ and $t^\prime\gg1$,
    \begin{equation}
        \langle v(t)v(t^\prime)\rangle \sim \exp\left[-\frac{2c}{1-\alpha}(t^{1-\alpha}-{t^\prime}^{1-\alpha})\right],
    \end{equation}
    which corresponds to Eq. (12) of the main text.

    \item[(ii) $\boldsymbol{\alpha>1}$ : Saturation to a Finite Limit]
    For $\alpha>1$, the sum $\sum_{u=t^\prime+1}^{t}u^{-\alpha}$ converges as $t\to \infty$. Therefore, with $t^\prime\gg1$ fixed, the velocity autocorrelation approaches a finite value,
    \begin{equation}
    \langle v(t)v(t^\prime)\rangle \quad \rightarrow \quad  \exp\left[-2c\sum_{u=t^\prime+1}^{\infty}u^{-\alpha}\right],
   \end{equation}
   which corresponds to Eq. (18) of the main text.
   
   Using the identity
   \begin{equation}
       \sum_{u=t^\prime+1}^{\infty}u^{-\alpha} = \zeta(\alpha)-\sum_{u=1}^{t^\prime}u^{-\alpha},
   \end{equation}
   this can equivalently be written as
   \begin{equation}
    \langle v(t)v(t^\prime)\rangle \sim \exp\left[-2c\left(\zeta(\alpha)-\sum_{u=1}^{t^\prime}u^{-\alpha}\right)\right].
\end{equation}

\item[(iii) $\boldsymbol{\alpha=1}$ : Algebraic Decay ]
At $\alpha=1$,
\begin{equation}
    \sum_{u=t^\prime+1}^{t}\frac{1}{u} \sim \ln\left(\frac{t}{t^\prime}\right),
\end{equation}
which gives
\begin{equation}
    \langle v(t)v(t^\prime)\rangle \sim \left(\frac{t^\prime}{t}\right)^{2c}, \quad t\to \infty,t^\prime\gg 1,
\end{equation}
This corresponds to Eq. (24) of the main text.
\end{description}

\section{MSD scaling from velocity position correlations}
\label{MSD_scaling}
In this section, we derive the asymptotic scaling of the mean square displacement (MSD) using the exact relation of MSD $\langle x^2(t)\rangle$ and the position velocity correlation $A(t)$, given in Eq. (5) of the main text, for the cases of power law, exponential, and linear flip probabilities.

\subsection{Long time asymptotic analysis for power law flip probability $\boldsymbol{p(t)=c/t^{\alpha}}$}
\label{MSD_scaling_power}
In this section, we analyze the long time asymptotic behavior of the position velocity correlation $A(t)$ and the mean square displacement $\langle x^2(t)\rangle$ for a persistent random walk with a power law flip probability $p(t)=c/t^{\alpha}$. The exact closed form expressions for $A(t)$ and $\langle x^2(t)\rangle$ are given in Eq. (5) of the main text. Here, we analyze three distinct asymptotic regimes determined by the value of the exponent $\alpha$.

\subsubsection{Anomalous diffusion regime ($\alpha<1$)}

Using the closed form expressions for $A(t)$ given in Eq. (5) of the main text, we write
\begin{equation}
\label{product_gamma}
    \prod_{u=s+1}^{t}\gamma(u) = \exp\left[\sum_{u=s+1}^{t} \ln\left(1-\frac{2c}{u^{\alpha}}\right)\right],
\end{equation}
we decompose the sum at a large but finite cutoff $U$($s\leq U<<t$):
\begin{equation}
 \sum_{u=s+1}^{t}\ln\left(1-\frac{2c}{u^{\alpha}}\right) =   \sum_{u=s+1}^{U}\ln\left(1-\frac{2c}{u^{\alpha}}\right)+ \sum_{u=U+1}^{t}\ln\left(1-\frac{2c}{u^{\alpha}}\right).
\end{equation}
The first term is a finite constant depending on $s$. Since for $\alpha<1$, the tail sum diverges with $t$, this constant does not affect the asymptotic behavior and may be neglected. For $u>>1$, we expand the logarithm to leading order,
\begin{equation}
   \ln\left(1-\frac{2c}{u^{\alpha}}\right) = -\frac{2c}{u^{\alpha}}+\mathcal{O}(u^{-2\alpha}),
\end{equation}
and retain only the dominant term. Restoring the lower limit to $s+1$ we obtain
\begin{equation}
\label{approx ln to leading order}
 \sum_{u=s+1}^{t}\ln\left(1-\frac{2c}{u^{\alpha}}\right)\approx -2c\sum_{u=s+1}^{t} u^{-\alpha}. 
\end{equation}
Approximating the sum by an integral yields 
\begin{equation}
\label{sum_by_integral}
   \sum_{u=s+1}^{t} u^{-\alpha} \approx \int_{s+1}^{t}u^{-\alpha}du = \frac{t^{1-\alpha}-(s+1)^{1-\alpha}}{1-\alpha},
\end{equation}
Substituting Eq. (\ref{sum_by_integral}) into Eq. (\ref{approx ln to leading order}), and then inserting the resulting expression into Eq. (\ref{product_gamma}), we obtain
\begin{equation}
\label{product_gamma_approx_into_integral}
\prod_{u=s+1}^{t}\gamma(u) \approx \exp\left[-\frac{2c}{1-\alpha}(t^{1-\alpha}-(s+1)^{1-\alpha})\right],
\end{equation}

Substituting Eq. (\ref{product_gamma_approx_into_integral}) into the exact expression of position velocity correlation $A(t)$ given in Eq. (\ref{A(t)}), we obtain
\begin{equation}
    A(t) = \sum_{s=1}^{t}\exp\left[-\frac{2c}{1-\alpha}(t^{1-\alpha}-(s+1)^{1-\alpha})\right].
\end{equation}
Since, the dominant contribution in the above sum arises from $s+1$ close to $t$, we write $s=t-r$ with $r<<t$ and expanding
\begin{equation}
    (s+1)^{1-\alpha} = t^{1-\alpha}-(1-\alpha)(r-1)t^{-\alpha} + \mathcal{O}((r-1)^2t^{-(1+\alpha)}),
\end{equation}
one obtains
\begin{equation}
    \prod_{u=s+1}^{t}\gamma(u)\approx \exp(-2c(r-1)t^{-\alpha}).
\end{equation}
The prefactor $\exp(2ct^{-\alpha})\to 1$ as $t\to \infty$, so it is irrelevant asymptotically. Hence, in the large time limit $t\to \infty$,
\begin{equation}
    A(t) \approx \sum_{r=0}^{t-1}\exp(-2crt^{-\alpha})=\frac{1-\exp(-2ct^{1-\alpha})}{1-\exp(-2ct^{-\alpha})}\approx \frac{t^{\alpha}}{2c},
\end{equation}
using $A(s)\xrightarrow[s\to\infty]{}s^{\alpha}/2c$ and $\gamma(s+1)=1-2c/(s+1)^{\alpha}$ in the exact MSD relation given in Eq. (5) of the main text, we obtain
\begin{equation}
    \gamma(s+1)A(s) \approx \left(1-\frac{2c}{(s+1)^{\alpha}}\right)\frac{s^{\alpha}}{2c}=\frac{s^{\alpha}}{2c}-\frac{s^{\alpha}}{(s+1)^{\alpha}}.
\end{equation}
For large $s$,
\begin{equation}
   \frac{s^{\alpha}}{(s+1)^{\alpha}} = \left(1+\frac{1}{s}\right)^{-\alpha} = 1-\frac{\alpha}{s}+ \mathcal{O}\left(\frac{1}{s^2}\right),
\end{equation}
so that
\begin{equation}
    \gamma(s+1)A(s) \approx \frac{s^{\alpha}}{2c}-1+\frac{\alpha}{s}+\mathcal{O}\left(\frac{1}{s^2}\right).
\end{equation}
Splitting the sum as $\sum_{s=0}^{t-1}=\sum_{s=0}^{S}+\sum_{s=S+1}^{t-1}$ with $S>>1$, the first part is finite. The leading contribution of the second sum is 
\begin{equation}
    \sum_{s=S+1}^{t-1}\frac{s^{\alpha}}{2c}\approx \frac{1}{2c}\int_{}^{t}dss^\alpha= \frac{t^{\alpha+1}}{2c(\alpha+1)}.
\end{equation}
This yields the asymptotic super-diffusive but sub-ballistic MSD scaling, $\langle x^2(t)\rangle \sim \frac{t^{\alpha+1}}{c(\alpha+1)}$ at $\alpha<1$ quoted in the main text [Eq. (14)].

\subsubsection{Marginal ballistic regime ($\alpha=1$)}

For $\alpha=1$, the logarithm of the persistence product reads 
\begin{equation}
    \ln\left(\prod_{u=s+1}^{t}\gamma(u)\right) = \sum_{u=s+1}^{t}\ln\left(1-\frac{2c}{u}\right).
\end{equation}
For large $u$, we expand the logarithm to leading order,
\begin{equation}
    \ln\left(1-\frac{2c}{u}\right) = -\frac{2c}{u} + \mathcal{O}(u^{-2}).
\end{equation}
Since $\sum_{u=s}^{t}u^{-1}\sim \ln(t/s)$ diverges logarithmically as $t\to\infty$, sub-leading corrections contribute only finite terms can be neglected. Approximating the sum by an integral, we obtain 
\begin{equation}
    \sum_{u=s+1}^{t}\ln\left(1-\frac{2c}{u}\right)\approx -2c\sum_{u=s+1}^{t}\frac{1}{u} \approx -2c\ln\left(\frac{t}{s+1}\right),
\end{equation}
which yields 
\begin{equation}
    \prod_{u=s+1}^{t}\gamma(u) \approx \left(\frac{s+1}{t}\right)^{2c},
\end{equation}

The position velocity correlation $A(t)$ given in Eq. (5) of the main text can then be approximated for large $t$ as
\begin{equation}
\label{A(t) scaling for alpha=1}
    A(t) \approx \sum_{s=1}^{t}\left(\frac{s+1}{t}\right)^{2c}\approx \frac{1}{t^{2c}}\int_{1}^{t}(y+1)^{2c}dy ,
\end{equation}
After evaluating the integral explicitly, $A(t)$ simplifies to
\begin{equation}
\label{exact form of A(t) for alpha = 1}
   A(t) \approx \frac{(t+1)^{2c+1}-2^{2c+1}}{(2c+1)t^{2c}}.
\end{equation}
In the asymptotic limit $t\to\infty$, the constant term $2^{2c+1}$ becomes negligible compared to $(t+1)^{2c+1}$, and one may write
\begin{equation}
    (t+1)^{2c+1} = t^{2c+1}\left(1+\frac{1}{t}\right)^{2c+1} \approx t^{2c+1}.
\end{equation}
Consequently, Eq. (\ref{exact form of A(t) for alpha = 1}) reduces asymptotically to
\begin{equation}
\label{A(t)_scaling_alpha=1}
    A(t) \sim \frac{t^{2c+1}}{(2c+1)t^{2c}} = \frac{t}{2c+1},
\end{equation}

Finally, inserting Eq. (\ref{A(t)_scaling_alpha=1}) into the exact MSD identity given in Eq. (5) of the main text and using 
\begin{equation}
    \gamma(s+1) = 1- \frac{2c}{s} + \mathcal{O}\left(\frac{1}{s^2}\right),\quad s\to \infty,
\end{equation}
we note that to leading order $\gamma(s+1)\to 1$. Retaining only the leading contribution in the MSD sum, we obtain
\begin{equation}
    \langle x^2(t)\rangle \sim \frac{1}{2c+1}t^2, \quad t\to \infty,
\end{equation}
This yields the asymptotic ballistic MSD scaling at $\alpha=1$ quoted in the main text [Eq. (26)].

\subsubsection{Ballistic regime with convergent persistence ($\alpha>1$)}

We start from the exact expression 
\begin{equation}
    \ln\left(\prod_{u=s+1}^{t}\gamma(u)\right) = \sum_{u=s+1}^{t}\ln\left(1-\frac{2c}{u^{\alpha}}\right).
\end{equation}
For $\alpha>1$, the series $\sum_{u=1}^{\infty}u^{-\alpha}$ converges. Therefore,
\begin{equation}
 \sum_{u=s+1}^{t}\ln\left(1-\frac{2c}{u^{\alpha}}\right) \xrightarrow[t\to\infty]{} \sum_{u=s+1}^{\infty}\ln\left(1-\frac{2c}{u^{\alpha}}\right)\equiv \ln c(s),
\end{equation} 
which defines a finite $s$ dependent constant $c(s)$. Hence,
\begin{equation}
    \prod_{u=s+1}^{t}\gamma(u) \xrightarrow[t\to\infty]{} c(s),
\end{equation}

To determine the contribution of large $s$ to $A(t)$, we examine the asymptotics of $c(s)$ 
for $s>>1$. Expanding the logarithm to leading order
\begin{equation}
    \ln\left(1-\frac{2c}{u^{\alpha}}\right) = -\frac{2c}{u^{\alpha}} + \mathcal{O}(u^{-2\alpha}).
\end{equation}
Thus,
\begin{equation}
    \ln c(s) \approx -2c\sum_{u=s+1}^{\infty} u^{-\alpha}.
\end{equation}
Using 
\begin{equation}
    \sum_{u=s+1}^{\infty}u^{-\alpha} = \zeta(\alpha)-\sum_{u=1}^{s}u^{-\alpha} \approx \zeta(\alpha)
    +\frac{s^{1-\alpha}}{\alpha-1} ,
\end{equation}
we obtain
\begin{equation}
    \ln c(s) \approx -2c\zeta(\alpha) - \frac{2c}{\alpha-1}s^{1-\alpha}, \quad s>>1.
\end{equation}
Exponentiating,
\begin{equation}
\label{c(s)}
 c(s) \approx C_0\exp\left[-\frac{2c}{\alpha-1}s^{1-\alpha}\right], \quad C_0 = e^{-2c\zeta(\alpha)}.  
\end{equation}
The position velocity correlation is 
\begin{equation}
    A(t) = \sum_{s=1}^{t}c(s).
\end{equation}
we decompose the sum as 
\begin{equation}
    A(t) = \sum_{s=1}^{S}c(s)+\sum_{s=S+1}^{t}c(s),
\end{equation}
where $S>>1$ is fixed.

The first sum contributes a constant. For large $s$, using Eq. (\ref{c(s)}) and expanding the exponential,
\begin{equation}
    c(s)\approx C_0\left[1-\frac{2c}{\alpha-1} s^{1-\alpha}+\mathcal{O}(s^{2-2\alpha})\right].
\end{equation}
Therefore,
\begin{equation}
    A(t) \approx C_0t-\frac{2cC_0}{\alpha-1}\sum_{s=1}^{t}s^{1-\alpha}+......
\end{equation}
Since, in the large time limit $t\to\infty$,
\begin{equation}
    \sum_{s=1}^{t}s^{1-\alpha} \approx \frac{t^{2-\alpha}}{2-\alpha}, \quad \alpha>1,
\end{equation}
We finally obtain
\begin{equation}
\label{A(t) for alpha>1}
    A(t)\approx C_0t-\frac{2cC_0}{(\alpha-1)({2-\alpha})}t^{2-\alpha}+.....
\end{equation}
The leading behavior is linear in $t$,
\begin{equation}
    A(t)\sim C_0t, \quad C_0=e^{-2c\zeta(\alpha)},
\end{equation}

Using the exact MSD identity given in Eq. (5) of the main text and noting that 
\begin{equation}
\label{gamma(s+1)_alpha>1}
    \gamma(s+1) = 1-\frac{2c}{s^{\alpha}}+\mathcal{O}(s^{-\alpha-1}), \quad s \to \infty,
\end{equation}
we observe that, to leading order, $\gamma(s+1)\to 1$. substituting Eq. (\ref{A(t) for alpha>1}) into Eq. (5) of the main text we obtain
\begin{equation}
    \langle x^2(t)\rangle \approx 2C_0\sum_{s=0}^{t-1}s-\frac{4cC_0}{(\alpha-1)({2-\alpha})}\sum_{s=0}^{t-1}s^{2-\alpha}+......
\end{equation}
Evaluating the sums
\begin{equation}
    \langle x^2(t)\rangle \approx C_0t^2-\frac{4cC_0}{(\alpha-1)({2-\alpha})}t^{3-\alpha}+....
\end{equation}
The leading behavior of MSD is therefore quadratic in $t$,
\begin{equation}
    \langle x^2(t)\rangle \sim C_0t^2, \quad C_0 = e^{-2c\zeta(\alpha)},
\end{equation}
This yields the asymptotic ballistic MSD scaling for all $\alpha>1$ quoted in the main text [Eq. (20)].

\subsection{Short and long time scaling for exponential flip probability $\boldsymbol{p(t)=p_0e^{-\lambda t}}$}
\label{MSD_scaling_expo}

In this section, we analyze the short and long time behavior of the position velocity correlation function $A(t)$ and the mean squared displacement $\langle x^2(t) \rangle$ for a persistent random walk with an exponentially decaying flip probability $p(t) = p_0e^{-\lambda t}$. The exponential protocol introduces a natural crossover timescale $t_c = \lambda^{-1}$, which separates an early time diffusive regime from a long time ballistic regime. Below we derive the scaling behavior in both limits.

\subsubsection{Short time diffusive regime ($t<<t_c=\lambda^{-1}$)}

The velocity persistence is 
\begin{equation}
    \gamma(t) = 1-2p_0e^{-\lambda t}
\end{equation}
For $t<<t_c=\lambda^{-1}$, Since $\lambda t<<1$, we expand the exponential as 
\begin{equation}
    e^{-\lambda t} = 1-\lambda t + \mathcal{O}(\lambda^2 t^2),
\end{equation}
which yields
\begin{equation}
    \gamma(t) = \gamma_0 + 2p_0 \lambda t + \mathcal{O}(\lambda^2 t^2), \quad \gamma_0 = 1-2p_0.
\end{equation}
The position velocity correlation function satisfies the exact recurrence relation (see Eq. (3) of the main text)
\begin{equation}
    A(t+1) = \gamma(t+1)A(t)+1, \quad A(0) = 0.
\end{equation}
Substituting the short time expansion of $\gamma(t)$, we obtain
\begin{equation}
    A(t+1) = \gamma_0 A(t) + 1 + 2p_0(\lambda t + \lambda)A(t) + \mathcal{O}(\lambda^2 t^2A(t)).
\end{equation}
At early times $A(t)$ remains finite, so the correction terms is of order $\lambda t<<1$. To leading order, the recurrence therefore reduces to 
\begin{equation}
    A(t+1) \approx \gamma_0 A(t) + 1, \quad t<<t_c.
\end{equation}
This recurrence is identical to that of a persistent random walk with constant persistence 
$\gamma_0$, and its solution is
\begin{equation}
    A(t) = \sum_{k=0}^{t-1}\gamma_0^k = \frac{1-\gamma_0^t}{1-\gamma_0}.
\end{equation}
For $t>>1$ but still $t<<t_c$, this yields
\begin{equation}
    A(t) \approx \frac{1}{1-\gamma_0}.
\end{equation}
Using the short time forms
\begin{equation}
\gamma(s+1) = \gamma_0 + \mathcal{O}(\lambda (s+1)), \quad A(s) = \frac{1}{1-\gamma_0} + \mathcal{O}(\lambda s),
\end{equation}
in the exact expression for the MSD  given in Eq. (5) of the main text, we find
\begin{equation}
    \gamma(s+1) A(s) = \frac{\gamma_0}{1-\gamma_0} + \mathcal{O}(\lambda s).
\end{equation}
Summing over $s$, the leading contribution is linear in time,
\begin{equation}
    \sum_{s=0}^{t-1}\gamma(s) A(s) = \frac{\gamma_0}{1-\gamma_0}t + \mathcal{O}(\lambda t^2).
\end{equation}
Since $\lambda t^2<<t$ in the regime $t<<t_c$, the correction term is sub-leading. Therefore the MSD scales as 
\begin{equation}
\label{MSD_exponential_flip_short_time}
    \langle x^2(t)\rangle \approx \left(\frac{1+\gamma_0}{1-\gamma_0}\right)t, \quad t<<t_c,
\end{equation}
demonstrating diffusive behavior at short times.

\subsubsection{Long time ballistic regime ($t>>t_c=\lambda^{-1}$)}

In this regime, We start from the exact expression 
\begin{equation}
    \ln\left(\prod_{u=s+1}^{t}\gamma(u)\right) = \sum_{u=s+1}^{t}\ln\left(1-2p_0e^{-\lambda u}\right).
\end{equation}
Since $\sum_{u=1}^{\infty}e^{-\lambda u}$ converges, the above sum converges as $t\to\infty$. We therefore define
\begin{equation}
 \sum_{u=s+1}^{t}\ln\left(1-2p_0e^{-\lambda u}\right) = \sum_{u=s+1}^{\infty}\ln\left(1-2p_0e^{-\lambda u}\right) \equiv \ln c(s),   
\end{equation}
which yields
\begin{equation}
    \prod_{u=s+1}^{t}\gamma(u) \xrightarrow[t\to\infty]{} c(s),
\end{equation}

For large $s$, we expand the logarithm to leading order,
\begin{equation}
    \ln(1-2p_0e^{-\lambda u}) = -2p_0e^{-\lambda u} + \mathcal{O}(e^{-2\lambda u}),
\end{equation}
which gives
\begin{equation}
    \ln c(s) \approx -2p_0\sum_{u=s+1}^{\infty}e^{-\lambda u}.
\end{equation}
Evaluating the geometric sum,
\begin{equation}
 \sum_{u=s}^{\infty}e^{-\lambda u} = \frac{e^{-\lambda (s+1)}}{1-e^{-\lambda}},   
\end{equation}
we obtain
\begin{equation}
    \ln c(s) \approx -\frac{2p_0e^{-\lambda (s+1)}}{1-e^{-\lambda}}, \quad s>>1.
\end{equation}
Exponentiating,
\begin{equation}
\label{c(s) for exponential flip prob}
    c(s) \sim \exp\left[-\frac{2p_0e^{-\lambda (s+1)}}{1-e^{-\lambda}}\right], \quad s>>1.
\end{equation}
The position velocity correlation is 
\begin{equation}
    A(t) = \sum_{s=1}^{t}c(s).
\end{equation}
we decompose the sum as 
\begin{equation}
    A(t) = \sum_{s=1}^{S}c(s)+\sum_{s=S+1}^{t}c(s),
\end{equation}
where $S>>1$ is fixed.

The first sum contributes a constant. For large $s$, expanding Eq. (\ref{c(s) for exponential flip prob}),
\begin{equation}
    c(s) \approx 1-\frac{2p_0e^{-\lambda (s+1)}}{1-e^{-\lambda}} + \mathcal{O}(e^{-2\lambda (s+1)}).
\end{equation}
Thus,
\begin{equation}
    A(t) \approx t-\frac{2p_0e^{-\lambda}}{1-e^{-\lambda}}\sum_{s=1}^{t}e^{-\lambda s}+....
\end{equation}
Since $\sum_{s=1}^{t}e^{-\lambda s}$ converges to a constant as $t\to \infty$, the correction is finite, and we obtain the leading behavior 
\begin{equation}
    A(t) \sim t, \quad t\to \infty,
\end{equation}

Using the exact MSD identity given in Eq. (5) of the main text and noting that $\gamma(s+1)=1-2p_0e^{-\lambda(s+1)}$ approaches unity exponentially fast as $s\to \infty$, we retain the leading order $\gamma(s+1)\approx 1$ and substitute $A(s)\sim s$ for large $s$ to obtain
\begin{equation}
    \langle x^2(t)\rangle \approx 2\sum_{s=0}^{t-1}s + \mathcal{O}(t) = t^2 + \mathcal{O}(t).
\end{equation}
Hence the leading behavior of MSD 
\begin{equation}
\label{MSD_exponential_flip_long_time}
    \langle x^2(t)\rangle \sim t^2, \quad t\to \infty,
\end{equation}
demonstrating ballistic scaling at long times for all $p_0>0$, as quoted in the main text.

\subsection{Short and long time scaling for linear flip probability $\boldsymbol{p(t)=p_0-\alpha t}$}
\label{MSD_scaling_linear}

In this section, we analyze the short and long time behavior of the position velocity correlation function $A(t)$ and the mean squared displacement $\langle x^2(t)\rangle$ for a persistent random walk with a linearly decreasing flip probability $p(t)=p_0-\alpha t$. This protocol introduces a finite cut off timescale $t_c = p_0/\alpha$, beyond which the flip probability vanishes identically. As a consequence, the dynamics exhibits a crossover from an early time diffusive regime to a long time ballistic regime. Below we derive the scaling behavior in both limits.

\subsubsection{Short time diffusive regime ($t<<t_c=p_0/\alpha$)}

For $t<<t_c$, we have $\alpha t<<p_0$, and the persistence parameter
\begin{equation}
    \gamma(t) = 1-2p(t) = \gamma_0+2\alpha t, \quad \gamma_0 = 1-2p_0.
\end{equation}
The position velocity correlation function satisfies the exact recurrence [see Eq. (3) of the main text]
\begin{equation}
    A(t+1) = \gamma(t+1)A(t)+1, \quad A(0) = 0.
\end{equation}
Substituting this short time form of $\gamma(t)$, we obtain
\begin{equation}
    A(t+1) = \gamma_0 A(t) + 1 + 2(\alpha t + \alpha)A(t).
\end{equation}
Since, $A(t)$ remains finite at early times and $\alpha t<<1$, the last term is sub-leading. To leading order, the recurrence reduces to
\begin{equation}
    A(t+1) \approx \gamma_0A(t) + 1.
\end{equation}
This recurrence is identical to that of a persistent random walk with constant persistence 
$\gamma_0$, and its solution is
\begin{equation}
    A(t) = \sum_{k=0}^{t-1}\gamma_0^k = \frac{1-\gamma_0^t}{1-\gamma_0}.
\end{equation}
For $t>>1$ but still $t<<t_c$, this yields
\begin{equation}
    A(t) \approx \frac{1}{1-\gamma_0}.
\end{equation}
Using $\gamma(s+1)\approx \gamma_0$ and the saturated form of $A(s)$, valid for $t<<t_c$, into the exact MSD relation given in Eq. (5) of the main text we find
\begin{equation}
\label{MSD_linear_flip_short_time}
 \langle x^2(t)\rangle \approx \left(\frac{1+\gamma_0}{1-\gamma_0}\right)t, \quad t<<t_c,
\end{equation}
demonstrating diffusive behavior at short times.

\subsubsection{Long time ballistic regime ($t>>t_c=p_0/\alpha$)}

For $t.\geq t_c$, the flip probability vanishes,
\begin{equation}
    p(t) = 0 \quad \Rightarrow \quad \gamma(t)=1,
\end{equation}
and the walker continues to move in the same direction thereafter.

The persistence product appearing in the definition of $A(t)$, $\prod_{u=s+1}^{t} \gamma(u)$, can be decomposed as 
\begin{equation}
    \prod_{u=s+1}^{t}\gamma(u) = \left(\prod_{u=s+1}^{t_c} \gamma(u)\right)\left(\prod_{u=t_c}^{t} 1\right) \equiv c(s),
\end{equation}
where $c(s)$ is a finite, t independent constant for fixed $s<t_c$ and $c(s)=1$ for $s>t_c$.

According the sum defining $A(t)$ can be split as
\begin{equation}
    A(t) = \sum_{s=1}^{t_c-1}c(s) + \sum_{s=t_c}^{t}1 = C+ (t-t_c+1),
\end{equation}
where $C=\sum_{s=1}^{t_c-1} c(s)$ is a constant. Thus for $t>>t_c$,
\begin{equation}
    A(t) \sim t.
\end{equation}
Substituting this linear growth into the MSD expression given in Eq. (5) of the main text and noting that $\gamma(s+1) = 1$ for $s\geq t_c$, we find the dominant contribution scales as 
\begin{equation}
\label{MSD_linear_flip_long_time}
    \langle x^2(t) \rangle \sim 2\sum_{s=t_c}^{t-1}s = (t-t_c)(t+t_c-1) \sim t^2, \quad t>>t_c.
\end{equation}
Thus, at long times the dynamics becomes ballistic.

\begin{figure}[h!]
    \includegraphics[width=0.7\linewidth]{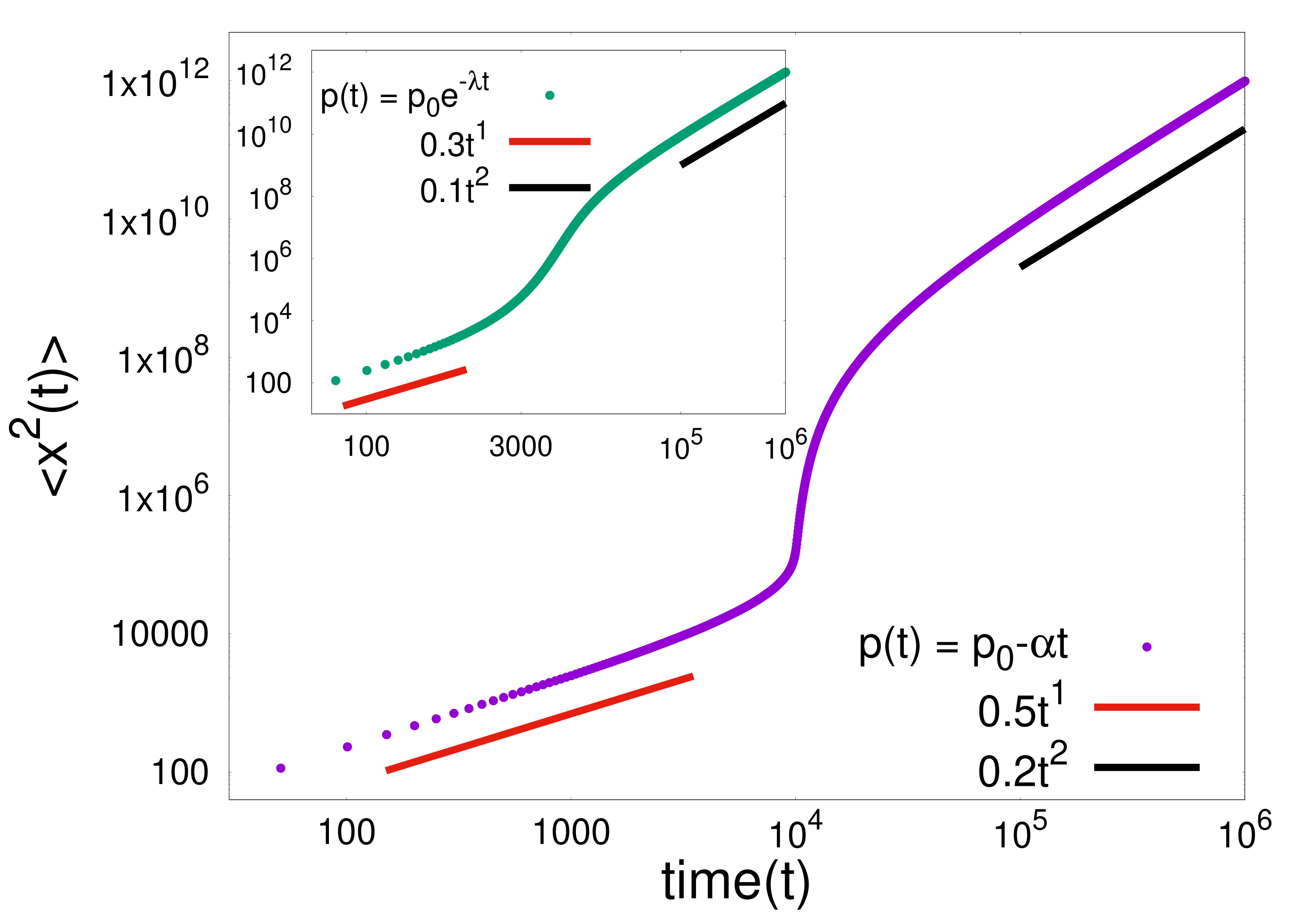}
    \caption{Mean-squared displacement $\langle x^2(t)\rangle$ shown on a log–log scale for time dependent flip probabilities. In the main panel, the MSD is shown for an linearly decreasing flip probability $p(t)=p_0-\alpha t$, where simulation data (points) are compared with analytical predictions (solid lines) [Eqs. (\ref{MSD_linear_flip_short_time}), (\ref{MSD_linear_flip_long_time})]. A crossover from diffusive behavior, $\langle x^2(t) \rangle \sim t$, at short times $t<<t_c=p_0/\alpha$ to ballistic growth, $\langle x^2(t)\rangle \sim t^2$, at long times $t>>t_c$ is clearly observed. The inset shows the corresponding MSD for a exponentially decaying flip probability $p(t)=p_0e^{-\lambda t}$, which also displays a transition from short time diffusion [Eq. (\ref{MSD_exponential_flip_short_time})] to long time ballistic motion [Eq. (\ref{MSD_exponential_flip_long_time})] at the cutoff time $t_c = 1/\lambda$.}
    \label{MSD vs t for exponential and linear flip}
\end{figure}

Figure \ref{MSD vs t for exponential and linear flip} shows the simulation results for exponentially decaying and linearly decreasing flip probabilities. In both cases, the dynamics exhibits a crossover from short-time diffusion to long-time ballistic growth at a characteristic timescale, consistent with the analytical predictions.

\section{Path Length Distribution $\boldsymbol{F(l)}$ }

We define a first passage event such that the walker changes direction i.e. tumbles for the first time since the start of the motion. Let $l$ denote the distance or number of steps traversed by the walker before this first flip occurs. We call $l$, the first persistence length or path length and denote its probability distribution by $F(l)$.

At time $t$, the probability that a walker flips its direction is assumed to decay as a power law
\begin{equation}
p(t) = ct^{-\alpha},
\end{equation}
where $\alpha>0$ is the memory exponent.

Consider a walker whose clock starts at time $t_0=1$. The probability that the walker remains in the same direction for at least $l$ consecutive steps i.e. has not yet flipped is 
\begin{equation}
P(l) = \quad \text{Pr(no flips for steps 1,2,3,....l)} = \prod_{t=1}^{l}[1-ct^{-\alpha}].
\end{equation}
Then the probability that the first flip occurs exactly at step $l$ is
\begin{equation}
F(l) = P(l-1)P_{flip}(l)=\left(\prod_{t=1}^{l}[1-ct^{-\alpha}]\right) \left(\frac{c}{l^{\alpha}}\right).
\end{equation}
Below we analyze the asymptotic forms of $P(l)$ and $F(l)$ in three distinct regimes of $\alpha$.

\begin{figure}[h!]
    \centering
    \includegraphics[width=0.7\linewidth]{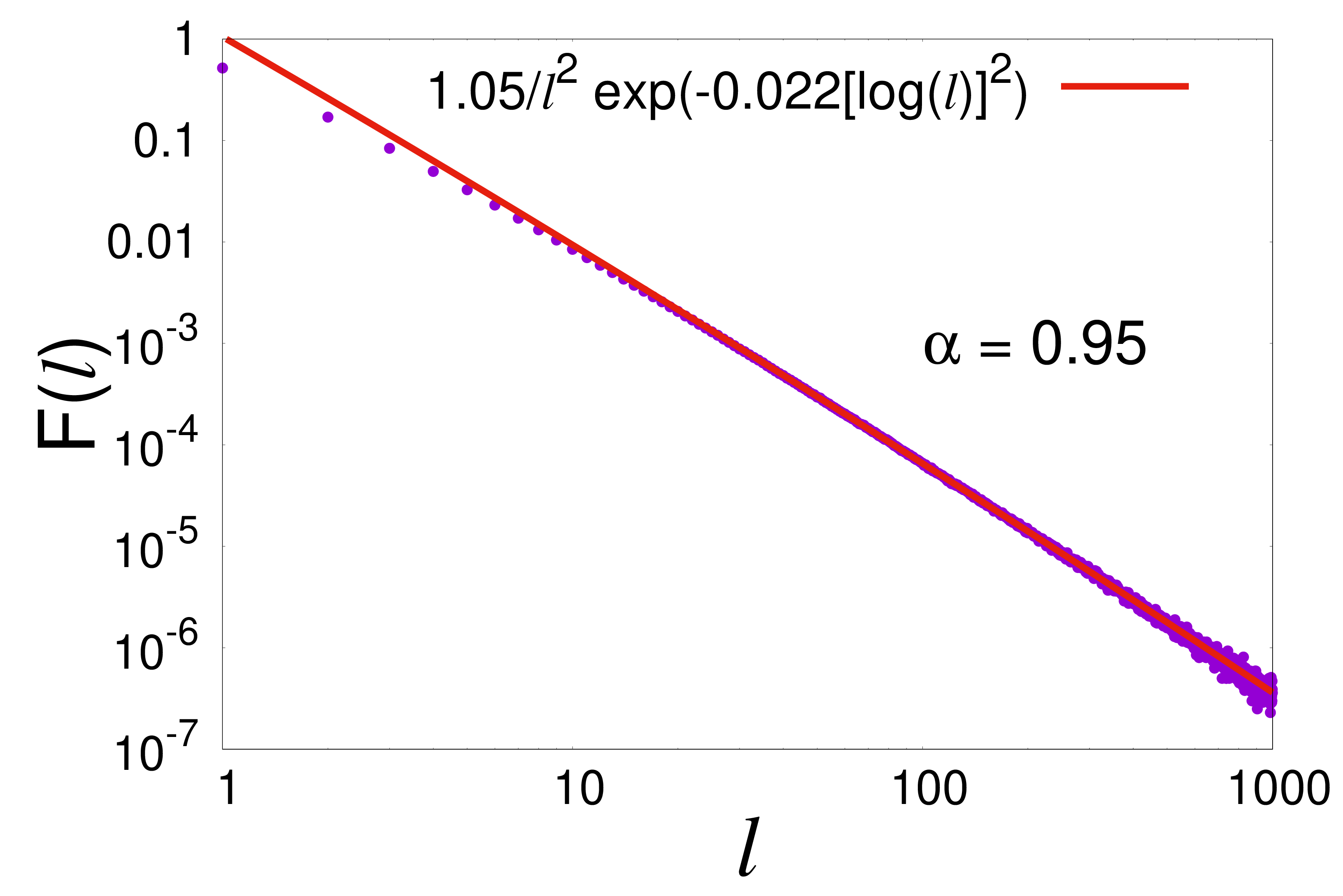}
    \caption{Plot of path length distribution $F(l)$ vs $l$ in log-log scale for $\alpha=0.95$. The data are well fitted by the form $F(l)\sim l^{-2}\exp\left[-D(\ln l)^2\right]$ with $D\approx0.022$ which is in excellent agreement with the analytical prediction for $\alpha$ very close to $1$ from below. [Eq. (\ref{form of F(l) with alpha very close to 1})]. simulations are performed with $c=1$.}
    \label{path length dist for alpha = 0.95}
\end{figure}

\begin{description}
    \item[Case I : $\boldsymbol{\alpha<1}$] 

To study the large $l$ behavior of $P(l)$ and $F(l)$, we take logarithms and expand 
\begin{equation}
\ln P(l) = \sum_{t=1}^{l} \ln(1-ct^{-\alpha}).
\end{equation}
For large $t$, $t^{-\alpha}<<1$, so
\begin{equation}
\ln (1-ct^{-\alpha}) \underline{\sim} -ct^{-\alpha} -\frac{1}{2}c^2 t^{-2\alpha} - .....
\end{equation}
Keeping only the leading term gives 
\begin{equation}
\label{ln P(l)}
 \ln P(l) \approx -c\sum_{t=1}^{l} t^{-\alpha} .  
\end{equation}
Approximating the sum by an integral for large $l$ in leading order gives 
\begin{equation}
\label{approximating sum by integral}
 \sum_{t=1}^{l} t^{-\alpha} \underline{\sim} \int_{1}^{l} x^{-\alpha} dx = \frac{l^{1-\alpha}-1}{1-\alpha}. 
\end{equation}
Thus,
\begin{equation}
 P(l) \sim \exp\left[-\frac{c}{1-\alpha}l^{1-\alpha}\right]  . 
\end{equation}
Hence, the first passage distribution becomes 
\begin{equation}
\label{tail of F(l)}
 F(l) = P(l-1)\left(\frac{c}{l^{\alpha}}\right) \sim \left(\frac{c}{l^{\alpha}}\right)  \exp\left[-\frac{c}{1-\alpha}l^{1-\alpha}\right].  
\end{equation}
These correspond to the forms of $P(l)$ and $F(l)$ quoted in Eqs. (16) and (17) of the main text.
This represents a stretched exponential tail of $P(l)$ and $F(l)$ (inset (a) of Fig. \ref{path length dist for alpha = 0.6,1,1.5}) and therefore all moments of $F(l)$ are finite.

Although the exact asymptotic form of $F(l)$ for $\alpha<1$ is stretched exponential [Eq. (\ref{tail of F(l)})], but that ceases to hold when $\alpha$ approaches unity from below.

In the regime $\alpha=1-\delta$ with $0<\delta<<1$, the exponent $l^{1-\alpha}=l^{\delta}$ varies extremely slowly with $l$, and the stretched exponential crossover gives way to a nearly power law behavior with a logarithmic correction [Fig. \ref{path length dist for alpha = 0.95}]. To capture this limit analytically we expand the survival probability in different orders of $\delta$. From Eq. (\ref{approximating sum by integral}) and using $1-\alpha=\delta$ we obtain
\begin{equation}
\sum_{t=1}^{l} t^{-\alpha} \underline{\sim} \int_{1}^{l} x^{-(1-\delta)} dx = \frac{l^{\delta}-1}{\delta}.
\end{equation}
Expanding $l^{\delta}=e^{\delta \ln l}\underline{\sim} 1 + \delta \ln l + \frac{\delta^2}{2}(\ln l)^2+...$,
\begin{equation}
  \frac{l^{\delta}-1}{\delta} = \ln l + \frac{\delta}{2} (\ln l)^2 + \mathcal{O} (\delta^2\ln^3l). 
\end{equation}
Thus,
\begin{equation}
P(l) \underline{\sim} l^{-c}\exp\left[-\frac{c\delta}{2}(\ln l)^2\right].
\end{equation}
So the corresponding path length distribution $F(l)$ for large $l$ becomes
\begin{equation}
\label{form of F(l) with alpha very close to 1}
  F(l) \underline{\sim} cl^{-(c+1)}\exp\left[-\frac{c(1-\alpha)}{2}(\ln l)^2\right], 
\end{equation}
with $\delta=1-\alpha<<1$.
\end{description}

\begin{figure}[h!]
    \centering
    \includegraphics[width=0.7\linewidth]{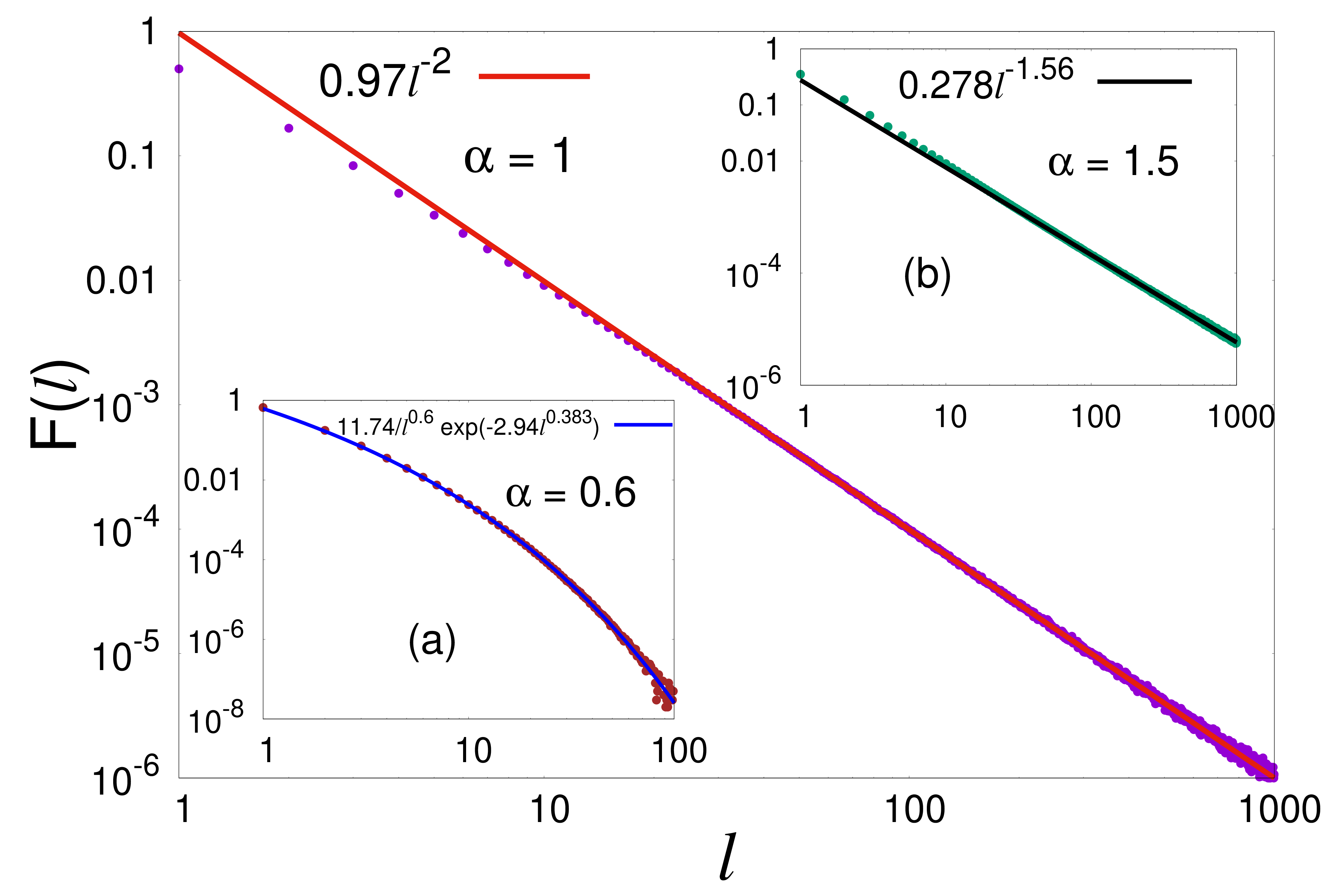}
    \caption{Plot of path length distribution $F(l)$ vs $l$ in log-log scale for $\alpha=1$. The tail of $F(l)$ follows a  power law of the form $\sim 1/l^{2}$, consistent with Eq. (\ref{F(l)_alpha=1}). Inset (a) corresponds to the plot of $F(l)$ vs $l$ in log-log scale for $\alpha=0.6$ where the tail of $F(l)$ follows $F(l)\sim l^{-0.6} \exp(-al^{b})$ with $a\approx2.94,b\approx0.383$, consistent with Eq. (\ref{tail of F(l)}). Inset (b) shows the plot of $F(l)$ vs $l$ in log-log scale for $\alpha=1.5$ where the tail of $F(l)$ remains a power law of the form $\sim 1/l^{\alpha}$ with $\alpha\approx 1.56$, consistent with Eq. (\ref{F(l)_alpha>1}). All simulations are performed with $c=1$. }
    \label{path length dist for alpha = 0.6,1,1.5}
\end{figure}

\begin{description}
    \item[Case II : $\boldsymbol{\alpha=1}$] 

At exactly $\alpha = 1$,
\begin{equation}
\sum_{t=1}^{l} t^{-1} \underline{\sim} \ln l + \gamma
\end{equation}
where $\gamma$ is the Euler–Mascheroni constant. Therefore,
\begin{equation}
P(l) \sim B l^{-c} \quad \text{with}\quad B = e^{-c\gamma}
\end{equation}
Using $F(l) = P(l-1)P_{flip}(l)$ and $P_{flip}(l)=cl^{-1}$, we obtain
\begin{equation}
\label{F(l)_alpha=1}
 F(l) \sim B^{\prime} l^{-(1+c)} \quad \text{with} \quad B^\prime = cB  
\end{equation}
These correspond to the forms of $P(l)$ and $F(l)$ quoted in Eq. (27) of the main text. The path length distribution at $\alpha=1$ is thus a pure power law with power law exponent $2$ [Fig. \ref{path length dist for alpha = 0.6,1,1.5}]. This implies that the mean path length which is the first moment of the path length distribution $F(l)$ diverges logarithmically at $\alpha=1$. 

\item[Case III : $\boldsymbol{\alpha>1}$] 

For $\alpha>1$, the series $\sum_{t=1}^{\infty} t^{-\alpha}$ converges to the Riemann zeta function $\zeta (\alpha)$. We define the asymptotic survival probability
\begin{equation}
    P_{\infty}(\alpha) = \lim_{l\to \infty}P(l),
\end{equation}
where $P(l)$ denotes the probability that the walker has not flipped its velocity up to time $l$. Taking the limit $l\to \infty$ in Eq. (\ref{ln P(l)}), we obtain
\begin{equation}
  \ln P_{\infty}(\alpha) = - c\sum_{t=1}^{\infty} t^{-\alpha} = -c\zeta (\alpha).
\end{equation}
So,
\begin{equation}
 P_{\infty}(\alpha) = e^{-c\zeta(\alpha)}  > 0.
\end{equation}
Thus the probability that the walker remain unflipped forever which is the survival probability, tends to a non zero constant at $\alpha>1$. For large but finite $l$,
\begin{equation}
\label{F(l)_alpha>1}
    F(l) = P(l-1)\left(\frac{c}{l^{\alpha}}\right) \sim cP_{\infty}(\alpha) l^{-\alpha}
\end{equation}
These correspond to the forms of $P_\infty(\alpha)$ and $F(l)$ quoted in Eq. (21) and (23) of the main text. In this regime at $\alpha>1$ the tail of first passage density $F(l)$ is still a power law with exponent $\alpha$ (inset (b) of Fig. \ref{path length dist for alpha = 0.6,1,1.5}). The existence of a non zero survivor fraction $P_{\infty}(\alpha)$ in this regime, implies that $F(l)$  is not normalized, with
\begin{equation}
    \sum_{l=1}^{\infty}F(l) = 1-P_{\infty}.
\end{equation}
The complete first passage description is therefore given by $\{F(l),P_{\infty}(\alpha)\}$. The first moment of this full distribution, corresponding to the mean path (persistence) length, still diverges at $\alpha>1$, similar to the marginal case $\alpha=1$, where the divergence is logarithmic. 
\end{description}

\section{Scaling of the displacement fluctuations $\boldsymbol{\sigma_x^2}$ near $\boldsymbol{\alpha=1}$}

In this section we derive the scaling form of the fluctuation of the displacement close to the critical point $\alpha=1$ for the persistent random walk with power law memory $p(t)=c/t^{\alpha}$. The displacement fluctuation is defined as 
\begin{equation}
    \sigma_x^2(t) = \langle x^2(t) \rangle - \langle x(t)\rangle ^2,
\end{equation}
which can be written in terms of velocity correlations as
\begin{equation}
    \sigma_x^2(t) = \sum_{u=1}^{t}\sum_{u^\prime=1}^{t}\langle v(u)v(u^\prime)\rangle.
\end{equation}
Using the symmetry of the correlation function this becomes
\begin{equation}
    \sigma_x^2(t) = t + 2\sum_{u=1}^{t}\sum_{u^\prime=1}^{u-1} \langle v(u)v(u^\prime)\rangle.
\end{equation}
For the present model the velocity correlations admit the exact representation [Eq. (6) of the main text]
\begin{equation}
    \langle v(u)v(u^\prime)\rangle = \prod_{k=u^\prime+1}^{u}\gamma(k), \quad u^\prime<u,
\end{equation}
where $\gamma(k)=1-2c/k^{\alpha}$.

Thus the displacement fluctuations can be written as 
\begin{equation}
    \sigma_x^2(t) = t + 2\sum_{u=1}^{t}\sum_{u^\prime=1}^{u-1}\prod_{k=u^\prime+1}^{u}\left(1-\frac{2c}{k^{\alpha}}\right).
\end{equation}
Introducing the separation $r=u-u^\prime$, we obtain
\begin{equation}
\label{sigma_x^2}
    \sigma_x^2(t) = t+2\sum_{u=1}^{t}\sum_{r=1}^{u-1}P(u,r),
\end{equation}
where
\begin{equation}
    P(u,r)  = \prod_{k=u-r+1}^{u}\left(1-\frac{2c}{k^{\alpha}}\right).
\end{equation}
These expressions correspond exactly to Eq. (28) of the main text.

To determine the long time behavior we analyze the dominant contribution to the double sum in the limit $t\to\infty$. The leading contribution arises from the region where $u\gg1$ and $u-r\gg1$, so that all terms in the product involve large $k$. In this regime the expansion $\ln(1-x)\approx -x$ for $x<<1$ becomes valid. The remaining regions of the sum, corresponding to small $u$ or 
small $u-r$, contribute only subleading corrections and therefore do not effect the asymptotic scaling behavior of $\sigma_x^2$.

Taking logarithms we obtain
\begin{equation}
    \ln P(u,r) = \sum_{k=u-r+1}^{u} \ln \left(1-\frac{2c}{k^{\alpha}}\right).
\end{equation}
For large $k$ the logarithm can be expanded to leading order, giving
\begin{equation}
    \ln P(u,r) \approx -2c \sum_{k=u-r+1}^{u}k^{-\alpha}.
\end{equation}
Near the critical point we write $\alpha=1+\epsilon$, with $|\epsilon|\ll1$. Replacing the sum by an integral gives
\begin{equation}
    \ln P(u,r) \approx -2c \int_{u-r}^{u}k^{-(1+\epsilon)}dk.
\end{equation}
evaluating the integral yields 
\begin{equation}
    \ln P(u,r) \approx \frac{2c}{\epsilon}\left[u^{-\epsilon}-(u-r)^{-\epsilon}\right].
\end{equation}
Expanding for small $\epsilon$, using $x^{-\epsilon}=e^{-\epsilon \ln x}$, and keeping terms up to first order in $\epsilon$, we obtain
\begin{equation}
    \ln P(u,r) \approx -2c \ln \frac{u}{u-r} + c\epsilon\left[(\ln u)^2-(\ln(u-r))^2\right].
\end{equation}
Thus
\begin{equation}
    P(u,r) \approx \left(\frac{u-r}{u}\right)^{2c} \exp[c\epsilon\{(\ln u)^2-(\ln(u-r))^2\}].
\end{equation}
Introducing $s=u-r$ and defining $\tau=s/u$, the sum over $s$ can be approximated by an integral for large $u$. Using $s=\tau u$ and $\ln s = \ln \tau + \ln u$, the exponent simplifies to
\begin{equation}
    (\ln u)^2- (\ln s)^2 = -2\ln \tau\ln u - (\ln \tau)^2.
\end{equation}
This gives
\begin{equation}
    P(u,\tau u) \approx \tau^{2c}\exp[-c\epsilon(\ln \tau)^2]\exp[-2c\epsilon \ln \tau \ln u].
\end{equation}
Substituting this expression into Eq. (\ref{sigma_x^2}) and interchanging the order of summation and integration yields 
\begin{equation}
    \sigma_x^2(t) \sim 2\int_{0}^{1}\tau^{2c}e^{-c\epsilon(\ln \tau)^2}\left[\sum_{u=1}^{t}u^{1-2c\epsilon \ln \tau}\right] d\tau.
\end{equation}
For large $t$, the sum behaves as 
\begin{equation}
    \sum_{u=1}^{t}u^p = \frac{t^{p+1}}{p+1}.
\end{equation}
With $p=1-2c\epsilon \ln \tau$, this leads to
\begin{equation}
    \sigma_x^2(t) \sim t^2 \int_{0}^{1}\frac{\tau^{2c}e^{-c\epsilon (\ln \tau)^2}e^{-2c\epsilon \ln \tau \ln t}}{1-c\epsilon \ln \tau}d\tau.
\end{equation}
Since $\epsilon=\alpha-1$, the displacement fluctuations $\sigma_x^2$ takes the scaling form
\begin{equation}
    \sigma_x^2(t)\sim t^2 \mathcal{G}[(\alpha-1)\ln t],
\end{equation}
where the scaling function is given by
\begin{equation}
 \mathcal{G}[(\alpha-1)\ln t] =  \int_{0}^{1}\frac{\tau^{2c}e^{-c(\alpha-1) (\ln \tau)^2}e^{-2c(\alpha-1) \ln \tau \ln t}}{1-c(\alpha-1) \ln \tau}d\tau.  
\end{equation}
These expressions correspond exactly to Eq. (29) and (30) of the main text.

\begin{figure}[h!]
    \centering
    \includegraphics[width=0.75\linewidth]{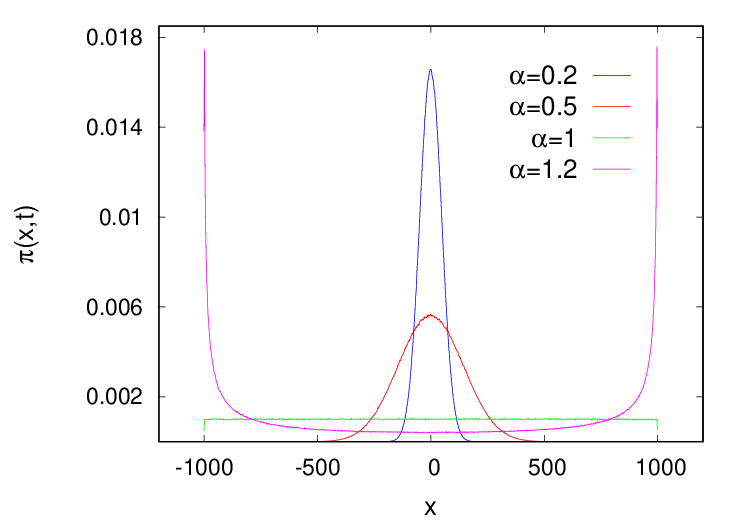}
    \caption{Distribution of the displacement of the walker is shown at time $t=1000$ for several values of $\alpha$. }
    \label{distribution}
\end{figure}

\begin{thebibliography}{99}

\bibitem{renshaw}
E. Renshaw and R. HENDER, {\em The correlated random walk}, J. Appl. Prob. {\bf 18}, 403 (1981).

\bibitem{Hanneken}
J. W. Hanneken, D. R. Franceschetti, {\em Exact distribution function for discrete time correlated random walks in one dimension}, J. Chem. Phys. {\bf 109}, 6533–6539 (1998).

\bibitem{weiss}
G. H. Weiss, {\em Some applications of persistent random walks and the telegrapher's equation}, Physica A {\bf 311}, 381 (2002). 

\bibitem{Kac}
M. Kac, {\em A Stochastic Model Related to the Telegrapher’s Equation}, Rocky Mountain Journal of Mathematics {\bf 4}, 497–509 (1974).

\bibitem{rosetto}
V. Rossetto, {\em The one-dimensional asymmetric persistent random walk}, J. Stat. Mech. 043204 (2018). 

\bibitem{Sadjadi}
Z. Sadjadi, M. Miri, {\em Persistent random walk on a one-dimensional lattice with random asymmetric transmittances}, Phys. Rev. E {\bf 78}, 061114 (2008).

\bibitem{Svenkeson}
A. Svenkeson, B. J. West, {\em Persistent random motion with maximally correlated fluctuations}, Phys. Rev. E {\bf 100}, 022119 (2019).

\bibitem{Marie}
M. Chupeau, O. B\'{e}nichou, R. Voituriez, {\em Mean cover time of one-dimensional persistent random walks}, Phys. Rev. E {\bf 89}, 062129 (2014). 

\bibitem{andre}
A. L. P. Livorati, T. Kroetz,C. P. Dettmann, I. L. Caldas, and E. D. Leonel, {\em Transition from normal to ballistic diffusion in a one-dimensional impact system}, Phys. Rev. E. {\bf 97}, 032205 (2018).

\bibitem{cenac}
P. C\'{e}nac, A. Le Ny, B. de Loynes and Y. Offret,{\em Persistent random walks. I. Recurrence versus transience},
J. Theor. Probab. {\bf 31(1)}, 232  (2018). 

\bibitem{Cenac}
P. C\'{e}nac, A. Le Ny, B. de L. and Y. Offret,  {\em Persistent Random Walks. II. Functional Scaling Limits}, J. Theor. Prob., {\bf 32(2)}, 633 (2019).

\bibitem{Masoliver}
J. Masoliver, J. M. Porra, G. H. Weiss, {\em Solutions of the telegrapher’s equation in the presence of traps}, Phys. Rev. A {\bf 45}, 2222 (1992).

\bibitem{heiko}
Z. Sadjadi, M. R. Shaebani and H. Rieger, and L. Santen, {\em Persistent random walk approach to anomalous transport of self-propelled particles}, Phys. Rev. E {\bf 91}, 062715 (2015).

\bibitem{Berg}
H. C. Berg, {\em Random Walks in Biology}, Princeton University Press, 1993.

\bibitem{berg}
H. C. Berg, D. A. Brown, {\em Chemotaxis in Escherichia coli analysed by Three-dimensional Tracking}, Nature {\bf 239}, 500–504 (1972).

\bibitem{Tailleur}
J. Tailleur, M. E. Cates, {\em Statistical Mechanics of Interacting Run-and-Tumble Bacteria}, Phys. Rev. Lett. {\bf 100}, 218103 (2008).

\bibitem{Benichou}
O. B\'{e}nichou, C. Loverdo, M. Moreau, and R. Voituriez, {\em Intermittent search strategies}, Rev. Mod. Phys. {\bf 83}, 81 (2011).

\bibitem{benichou}
O. B\'{e}nichou and R. Voituriez, {\em From first-passage times of random walks in confinement to geometry-controlled kinetics}, Phys. Rep. {\bf 539}, 225 (2014).

\bibitem{Marchetti}
M. C. Marchetti et al., {\em Hydrodynamics of soft active matter}, Rev. Mod. Phys. {\bf 85}, 1143 (2013).

\bibitem{Bechinger}
C. Bechinger et al., {\em Active particles in complex and crowded environments}, Rev. Mod. Phys. {\bf 88}, 045006 (2016).

\bibitem{Elgeti}
J. Elgeti, R. Winkler, and G. Gompper, {\em Physics of microswimmers—single particle motion and collective behavior: a review}, Rep. Prog. Phys. {\bf 78}, 056601 (2015).

\bibitem{Solon}
A. P. Solon et al., {\em Pressure and Phase Equilibria in Interacting Active Brownian Spheres}, Phys. Rev. Lett. {\bf 114}, 198301 (2015).

\bibitem{Zaburdaev}
V. Zaburdaev, S. Denisov, and J. Klafter, {\em L\'{e}vy walks}, Rev. Mod. Phys. {\bf 87}, 
483 (2015).

\bibitem{Metzler}
R. Metzler and J. Klafter, {\em The random walk's guide to anomalous diffusion: a fractional dynamics approach}, Phys. Rep. {\bf 339}, 1 (2000).

\bibitem{metzler}
R. Metzler et al., {\em Anomalous diffusion models and their properties: non-stationarity, non-ergodicity, and ageing at the centenary of single particle tracking }, Phys. Chem. Chem. Phys. {\bf 16}, 24128 (2014).

\bibitem{Barkai}
E. Barkai, Y. Garini, and R. Metzler, {\em Strange kinetics of single molecules in living cells}, Phys. Today {\bf 65}, 29 (2012).

\bibitem{Fedotov}
S. Fedotov, {\em Subdiffusion, chemotaxis, and anomalous aggregation }, Phys. Rev. E {\bf 83}, 021110 (2011).

\bibitem{barkai}
E. Barkai and Y. Cheng, {\em Aging continuous time random walks}, J. Chem. Phys. {\bf 118}, 6167 (2003).

\bibitem{Jeon}
J. H. Jeon and R. Metzler, {\em Fractional Brownian motion and motion governed by the fractional Langevin equation in confined geometries}, Phys. Rev. E {\bf 81}, 021103 (2010).

\bibitem{Schulz}
J. H. P. Schulz, E. Barkai, R. Metzler, {\em Aging Effects and Population Splitting in Single-Particle Trajectory Averages}, Phys. Rev. Lett. {\bf 110}, 020602 (2013).





\bibitem{SM}
See Supplemental Material for detailed additional
derivations along with other related
discussions.
\end{thebibliography}
\end{document}